\documentclass{article}

\usepackage{arxiv}

\usepackage[utf8]{inputenc} 
\usepackage[T1]{fontenc}    
\usepackage{hyperref}       
\usepackage{url}            
\usepackage{booktabs}       
\usepackage{amsfonts}       
\usepackage{nicefrac}       
\usepackage{microtype}      
\usepackage{lipsum}		
\usepackage{doi}
\usepackage{subcaption}
\usepackage{multirow}
\usepackage{wrapfig}
\usepackage{graphicx}
\usepackage[table]{xcolor}
\usepackage{floatrow}
\newfloatcommand{capbtabbox}{table}[][\FBwidth]
\usepackage{blindtext}

\title{The AeroSonicDB (YPAD-0523) dataset for acoustic detection and classification of aircraft}

\author{
    Blake Downward \\
        AeroSonicDB\\
	\And
    Jon Nordby \\
	Soundsensing AS \\
}

\date{}

\renewcommand{\headeright}{Technical Report}
\renewcommand{\undertitle}{Technical Report}
\renewcommand{\shorttitle}{\textit{arXiv} Template}

\begin{document}

\maketitle
\renewcommand{\abstractname}{\vspace{-\baselineskip}} 

\begin{abstract}	\noindent
The time and expense required to collect and label audio data has been a prohibitive factor in the availability of domain specific audio datasets. As the predictive specificity of a classifier depends on the specificity of the labels it is trained on, it follows that finely-labelled datasets are crucial for advances in machine learning. Aiming to stimulate progress in the field of machine listening, this paper introduces \textit{AeroSonicDB (YPAD-0523)}\footnote{\url{https://zenodo.org/records/8371595}\cite{downward_blake_2023_8371595}}, a dataset of low-flying aircraft sounds for training acoustic detection and classification systems. This paper describes the method of exploiting ADS-B radio transmissions to passively collect and label audio samples. Provides a summary of the collated dataset. Presents baseline results from three binary classification models, then discusses the limitations of the current dataset and its future potential.

The dataset contains 625 aircraft recordings ranging in event duration from 18 to 60 seconds, for a total of 8.87 hours of aircraft audio. These 625 samples feature 301 unique aircraft, each of which are supplied with 14 supplementary (non-acoustic) labels to describe the aircraft. The dataset also contains 3.52 hours of ambient background audio ("silence"), as a means to distinguish aircraft noise from other local environmental noises. Additionally, 6 hours of urban soundscape recordings (with aircraft annotations) are included as an ancillary method for evaluating model performance, and to provide a testing ground for real-time applications.

\end{abstract}

\quad\rule{425pt}{0.4pt}

\section{Introduction}
Excessive exposure to environmental noise can negatively impact the health and well-being of individuals. The World Health Organisation (WHO) ranks the burden of disease resulting from environmental noise as the second highest after air pollution. For aircraft specifically, the WHO strongly recommends reducing noise levels to below 45 dB \textit{L}\textsubscript{den}, as exposure to noise beyond this level has been shown to negatively impact health \cite{World_Health_Organisation_Regional_Office_for_Europe2019-za}.

To help understand and manage these risks, aircraft noise monitoring has been well established in many countries, leading to international standards for monitoring, detecting and classifying aircraft sound in the vicinity of airports \cite{ISO20906}. As urban populations expand and demand for air travel continues to grow, so too will the challenge of managing aircraft noise, and the need for ongoing development of noise monitoring and modelling systems.

In order to advance any field with machine learning, it is necessary to have well-annotated datasets suited to the task at hand. Although aircraft noise falls under the general task of environmental sound classification, generalised environmental sound classification datasets may not deliver the granularity required for practical deployment of an aircraft specific classifier. This paper presents a finely-labelled dataset of aircraft sounds, the inexpensive method to collect it, and a benchmark for model performance. All Python scripts and Jupyter Notebooks for downloading the dataset, extracting features, training models and reproducing the results presented in this paper are available on GitHub\footnote{\url{https://github.com/aerosonicdb/AeroSonicDB-YPAD0523}}.

\section{Background}

\subsection{Airport noise monitoring and modelling}

Airservices Australia operates a Noise and Flight Path Monitoring System (NFPMS) around Adelaide Airport \cite{adelaide_nfpms}. This system includes five Environmental (noise) Monitoring Units (EMU), four of which are located within 1km of the end of each runway, and the fifth is approximately 7km north-east of runway 23 in close proximity to the Adelaide CBD. These units are equipped with a microphone to continuously log environmental sound levels, and are interfaced with The Australian Advanced Air Traffic System (TAAATS) for aircraft positional information to assist correlating environmental noise events with overflying aircraft \cite{noise_fp_monitoring}. 

The purpose of this system is not to ensure compliance with aircraft noise regulations, as Australia does not have any regulations regarding a maximum allowable level of aircraft noise \cite{airservices_noise_monitoring}. Rather this system operates to inform community, governmental and industry stakeholders about the overall impact of current and proposed air operations for airports. 

Responsible for civilian aircraft noise complaints and enquiries, Airservices Australia can utilise the data from monitoring units to help verify, understand and resolve noise complaints. Additionally, data gathered from the NFMPS is used as a means to validate the accuracy of modelling which forecasts future noise impacts on the community given expected air traffic, flight paths and composition of the air fleet. 

\subsection{Automatic Dependent Surveillance-Broadcast}
ADS-B stands for "Automatic Dependent Surveillance-Broadcast" \cite{casa_ads-b_def} - a system which \textit{automatically broadcasts} an aircraft's \textit{GPS-dependent} data for \textit{surveillance} and situational awareness. ADS-B messages are digitally encoded radio signals transmitted at a frequency of 1090 MHz. Information transmitted via ADS-B includes positional data such as; latitude and longitude, speed, vertical rate, heading and altitude - along with the aircraft's unique identifier and a range of status messages. 

Since 2017, it has been mandatory for aircraft operating in Australia under Instrument Flight Rules (IFR) to be equipped to transmit ADS-B messages \cite{casa_cao_20.18}\cite{casa_vfr_ads-b}. This has led to virtually all aircraft operating in urban airspace to be equipped with ADS-B transmitters.

For this reason, aircraft monitoring has become more accessible to the general public in recent history. Popular flight tracking websites such as Flightradar24 and ADSB Exchange, aggregate ADS-B data from around the world to provide users with near real-time access to aircraft positional information. Aircraft can however be monitored locally without internet access. An inexpensive software defined radio (SDR) and an open-source digital decoder can effectively turn any PC into an ADS-B receiver via a USB port.

\subsection{Related work}
\subsubsection{Environmental Sound Datasets}
The general task of detecting potentially noisy sounds from the environment (\emph{Environmental Sound Classification}) is well established in the machine learning community. There are several large datasets dedicated to this task\cite{cartwright2020sonyc}\cite{salamon_dataset_2014:urbansound8k}, some of which include aircraft sounds\cite{piczak2015dataset:esc50}\cite{audio_set}.  Google's Audio Set is a large collection of human-labelled 10-second audio clips sourced from YouTube videos. Audio Set has a number of labels related to aircraft sounds, including; "Aircraft engine" (3.2 hrs), "Aircraft" (15.2 hrs),"Fixed-wing aircraft/airplane" (8.3 hrs),"Propeller/airscrew" (2.9 hrs) and "Jet engine" (2.1 hrs). The ESC-50 dataset contains samples for both "Airplane" and "Helicopter" - however with just 200 seconds of audio per class.

\subsubsection{Environmental Sound Classification}
 The attractiveness of inexpensive acoustic monitoring is being realised by researchers and developers in a wide range of applications. Acoustic sensors are being used to detect and timestamp gunshots near a shooting range. This is helping to deal with community complaints and ensure noisy activities do not exceed monthly limits\cite{nordby2021automatic}. On a larger scale, an NYU initiative called \textit{Sounds of New York City (SONYC)} is aiming to develop a systematic, city-wide approach for monitoring noise pollution and understanding which activities or sources contribute the most to the overall burden of noise\cite{cartwright2020sonyc}.

\subsubsection{Acoustic detection and classification of aircraft}
Aircraft noise has been well studied and featured in academic literature for a number of decades\cite{nasa-aircraft-noise-prediction}. More recently, advances in computing power which allow real-time inference of acoustic signals, have led to a number of researchers exploring the use of passive acoustic systems for detecting low-flying aircraft\cite{shi2011detecting}\cite{aircraft-detection-2009}\cite{acoustic-aircraft-tracking-2013}. In particular, Unmanned Aerial Vehicles (UAVs) have received a great deal of attention\cite{drone-dataset-2022}. Due to their low-altitude capabilities and small surface area, UAVs are difficult to detect with traditional radar, making them a popular choice for reconnaissance and/or deploying payloads to protected areas. Acoustic detection sensors/networks have been proposed as a means to counter these vulnerabilities and complement existing radar systems\cite{drone-detection-2021}\cite{c-uas-2021}.

The method for automatically capturing and labelling aircraft sounds is an adaptation of that described in \textit{Real-time identification of aircraft sound events}\cite{giladi2020real}.

\section{Methods}

\subsection{Recording locations}

\begin{figure}[ht]
\centering
\begin{subfigure}{0.5\textwidth}
    \centering
    \includegraphics[width=0.9\linewidth]{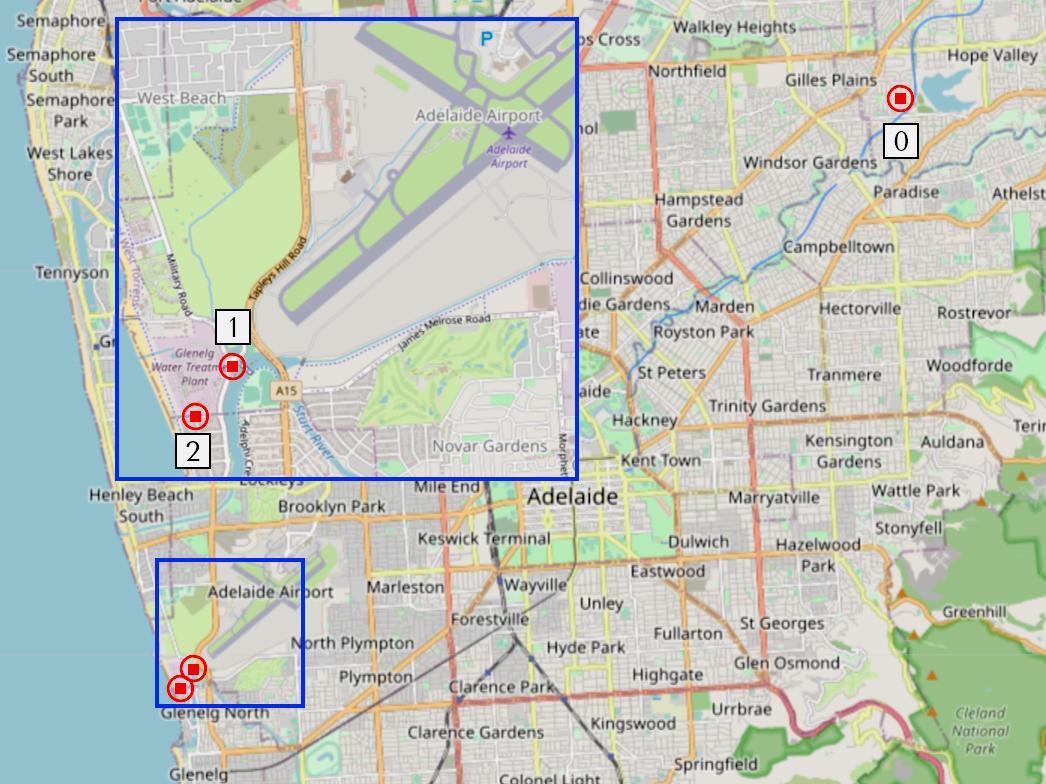}
    \caption{Recording locations}
    \label{fig:data-collect-center-locations}
\end{subfigure}%
\begin{subfigure}{0.5\textwidth}
    \centering
    \includegraphics[width=0.9\linewidth]{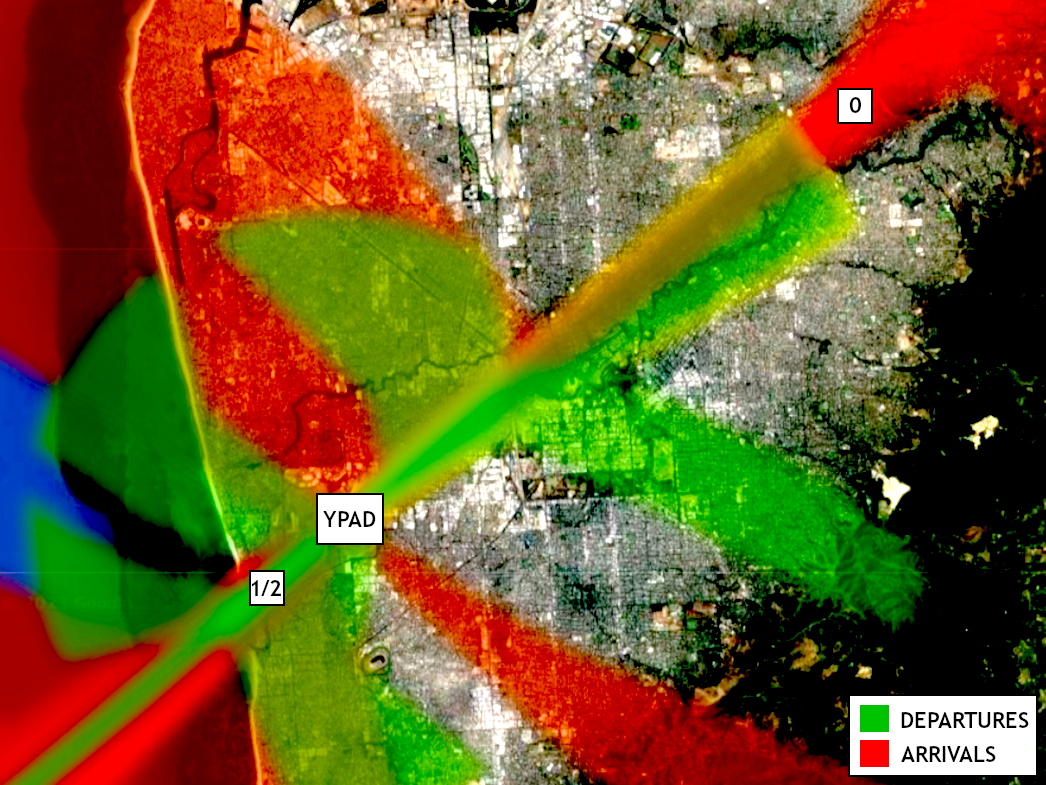}
    \caption{Heatmap of YPAD air traffic (June, 2023)}
    \label{fig:ypad-flight-paths}
\end{subfigure}
\caption{Locations of the recording devices across Adelaide. Location "0" is approximately 15.5km north-east of the airport. Locations "1" and "2" are approximately 500m and 1,000m south-west of the runway respectively. a) Image credit OpenStreetMap.org  b) Image credit: Airservices Australia. }
\label{fig:adelaide-map}
\end{figure}

Audio was recorded at three locations across the city of Adelaide, Australia (as shown in Figures \ref{fig:data-collect-center-locations} and \ref{fig:ypad-flight-paths}). Location "0" is approximately 15.5km north-east of runway "23" in a suburban setting. Aircraft at this location were recorded at a median altitude of 3,275 feet. In contrast, locations "1" and "2" are both situated within 1km south-west of runway "05", with the majority of aircraft captured at altitudes below 1,000 feet.

\subsection{Aircraft monitoring with ADS-B}

For this task, a \textit{Nooelec NESDR Mini 2+} SDR dongle\cite{nooelec_sdr} was used to receive digitally encoded radio transmissions at 1090 MHz.

With the aide of an open-source ADS-B decoder, \textit{Dump1090}\cite{dump_1090}, transmissions are decoded and piped to a custom Python script which then monitors the position and status of nearby aircraft. 

Messages broadcast from an aircraft include details such as; the aircraft's unique identifier (ICAO 24-bit address/hex code), GPS coordinates, altitude, heading and speed. By monitoring these messages, it is possible to track nearby aircraft and trigger a recording when they approach the device.

\begin{figure}[ht]
\centering
\includegraphics[width=0.6\textwidth]{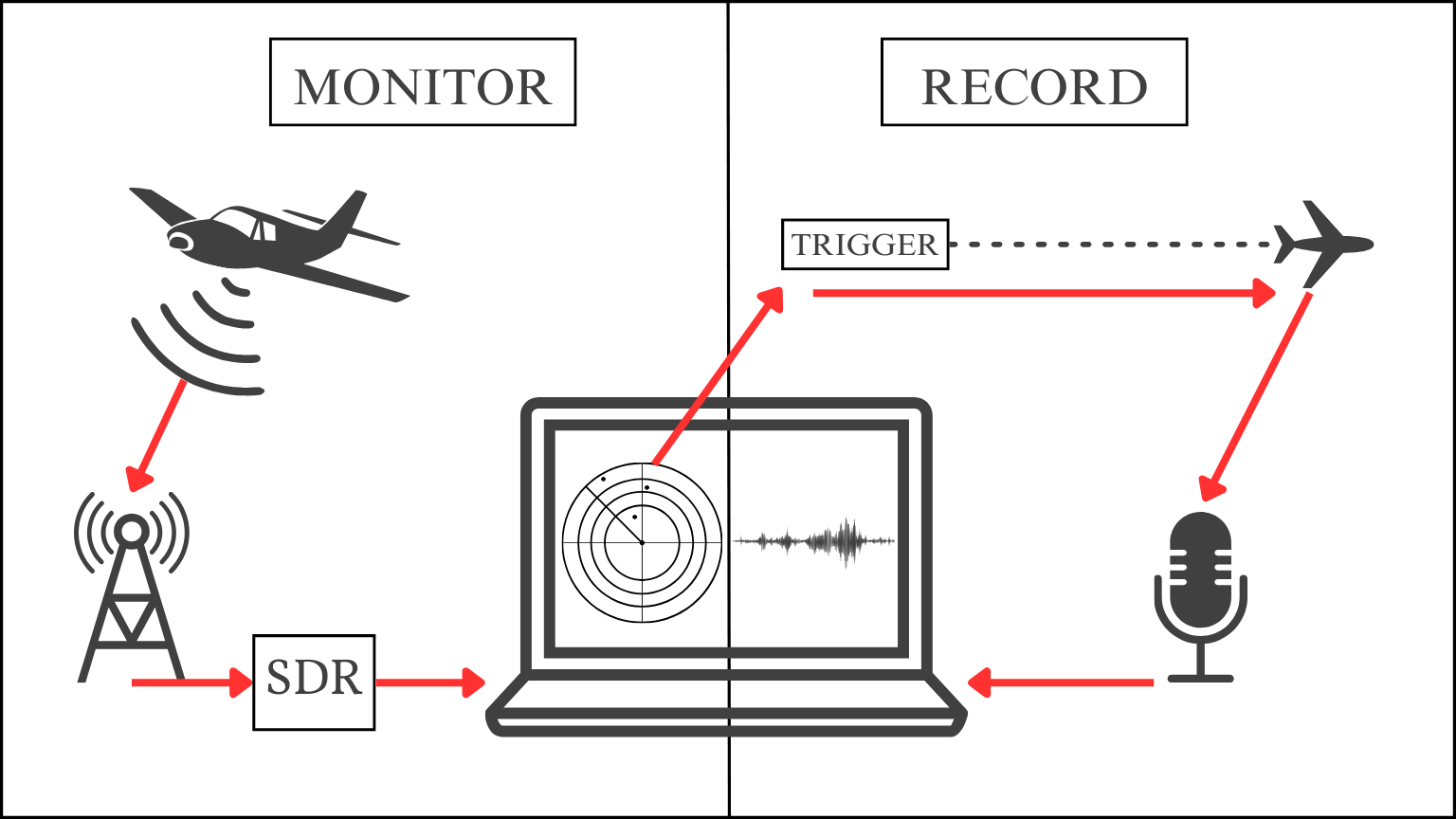}
\caption{LEFT: Flow of information from aircraft to SDR to device (monitoring). RIGHT: 60-second recording triggered by aircraft positional information.}
\label{fig:monitor-and-record}
\end{figure}

Once triggered, the aircraft's unique ICAO hex code and last reported altitude are included with the date, time, microphone and location as meta information in the filename. The flow of the monitoring and recording process is graphically demonstrated in Figure \ref{fig:monitor-and-record}.

\subsection{Audio recording setup}

\subsubsection{Microphone setup}
Two different microphone setups were used to record audio, these are identified in the dataset by the "mic" tag.

Microphone ID "0": Shure SM58 dynamic microphone with a Focusrite Scarlett Solo 2nd Generation audio interface.

Microphone ID "1": Samson Go Mic, USB condenser microphone (set to cardioid polar pattern).

Over the data collection period, various gain settings were used for both microphone setups, as such, microphone sensitivity cannot be assumed as a constant for each microphone or recording.

\subsubsection{Recording Trigger setup}
Two classes of audio recordings are triggered by the ADS-B monitoring setup described above; the negative class "0" (no aircraft/silence), and positive class "1" (aircraft).

A 60-second "aircraft" recording is triggered when an aircraft is within a specified distance of the recording device. This "trigger distance" varies between locations due to the difference in typical aircraft altitudes at each location. The trigger distance for location "0" is set to 3km, whereas locations "1" and "2" are 1km and 1.5km respectively.

Alternatively,  a 10-second "silence" recording is triggered only when ADS-B data consecutively confirms the absence of airborne aircraft within a 10km radius of the recording device. The close proximity of locations "1" and "2" to the airport made collecting silence samples more challenging than location "0". The ADS-B monitoring script only tracked airborne aircraft, not those transmitting an "on ground" message. This means there is a potential for "on ground" aircraft noise from the airport to leak into "silence" samples. Care was taken to ensure there was no salient aircraft noise present in the "silence" audio from these locations.

\subsubsection{File format and naming}
Audio files are recorded at a sample rate of 22,050 Hz and bit depth of 16-bit in WAV file format.

At the time of recording, files are tagged with their class, date, time, location ID and microphone ID. Additionally, aircraft recordings are tagged with the aircraft's ICAO 24-bit address (a globally unique identifier for aircraft) and last reported altitude before recording (altitude is removed from the filename once meta information has been stored).

General convention: "\{\textit{Class/Aircraft ID}\}\_\{\textit{Date}\}\_\{\textit{Time}\}\_\{\textit{Location ID}\}\_\{\textit{Microphone ID}\}.wav"

Aircraft example: "\textit{7C7CD0\_2023-05-09\_12-42-55\_2\_1.wav}"

Silence example: "\textit{000000\_2023-05-09\_12-30-55\_2\_1.wav}"

\subsection{Dataset preparation}
After audio samples are collected, they are individually human verified to ensure the recordings are of good quality and correctly labelled. Additionally for aircraft samples, the unique ICAO hex ID is used to retrieve further information about the airframe, such as manufacturer, model, engine type and various other features described in Table \ref{table:airframe-features}.

\subsubsection{Aircraft meta discovery}
The unique ICAO hex code transmitted via ADS-B does not directly hold information about the aircraft it identifies. In order to translate the hex code into relevant features of the aircraft, a number of publicly available databases \cite{adsb-exchange}\cite{radarbox}\cite{airport-data}\cite{plane-spotters} were cross-referenced to ensure a majority consensus for the tail-sign/registration number attributed to the hex code.

Next, the registration (tail-sign) obtained in the above step is used to search the Australian database of registered aircraft \cite{casa-aircraft-register}, which is regularly maintained and updated by the Australian civil aviation authority, CASA. Airframe features for Australian registered aircraft have been obtained from this database, however it does not contain information about internationally registered, or military designated aircraft. 

For international aircraft, a further round of discovery and consensus was required to achieve the same level of labelling as Australian registered aircraft. This included cross-referencing information from multiple sources, then ensuring there are no conflicts for that type of aircraft within the Australian database. All military aircraft were dropped from the dataset at this point. 

For a summary of the airframe features available in this dataset, see Table \ref{table:airframe-features}.

\subsubsection{Human verification and trimming}
\begin{figure}[ht]
\centering
\includegraphics[width=0.6\textwidth]{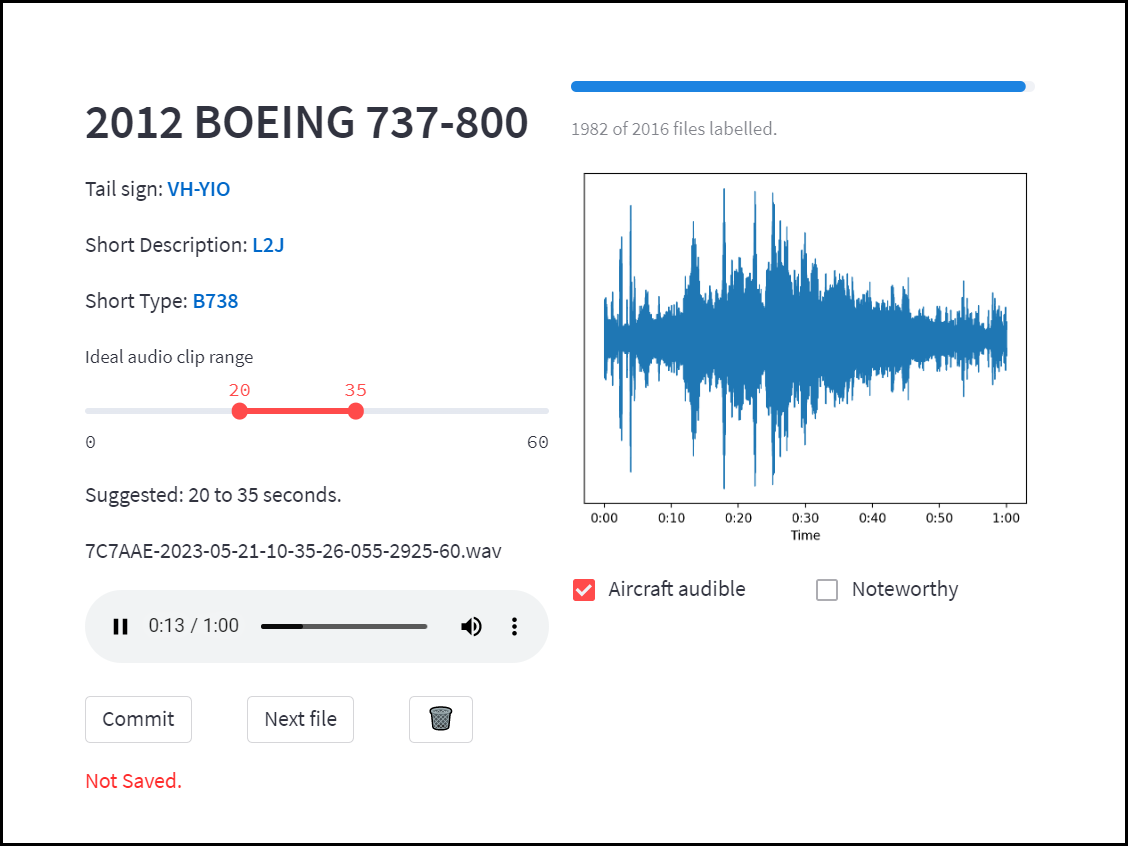}
\caption{Screenshot of the verification and trimming application.}
\label{fig:verify-and-trim-app}
\end{figure}

Audio samples and relevant meta information are each loaded into a custom application (Figure \ref{fig:verify-and-trim-app}) to assist with the human verification process. Samples are played to verify if an aircraft is audible or not. Samples of poor quality, or those which captured aircraft at an altitude greater than 10,000 feet were discarded at this point. Samples including identifiable human speech were either trimmed or dropped completely for privacy reasons. Finally, aircraft recordings are annotated with timestamps to indicate the start and end of the audio event. 

\subsubsection{Train, validation and test splits}
The dataset has been split into 6 approximately equal folds to ensure a common base for cross-validation. This includes 5 folds of training data for cross validation, and a single fold to be held-out as a test set for evaluation. To avoid data leakage between folds, samples were grouped by recording-session. Each recording session is defined by the location, microphone and time of the session. The test set only included samples that were collected at a time following the last recording session in the training set per location.

\subsubsection{Event/sample length}
\begin{figure}[ht]
\centering
\begin{subfigure}{0.5\textwidth}
    \centering
    \includegraphics[width=0.95\linewidth, height=6cm]{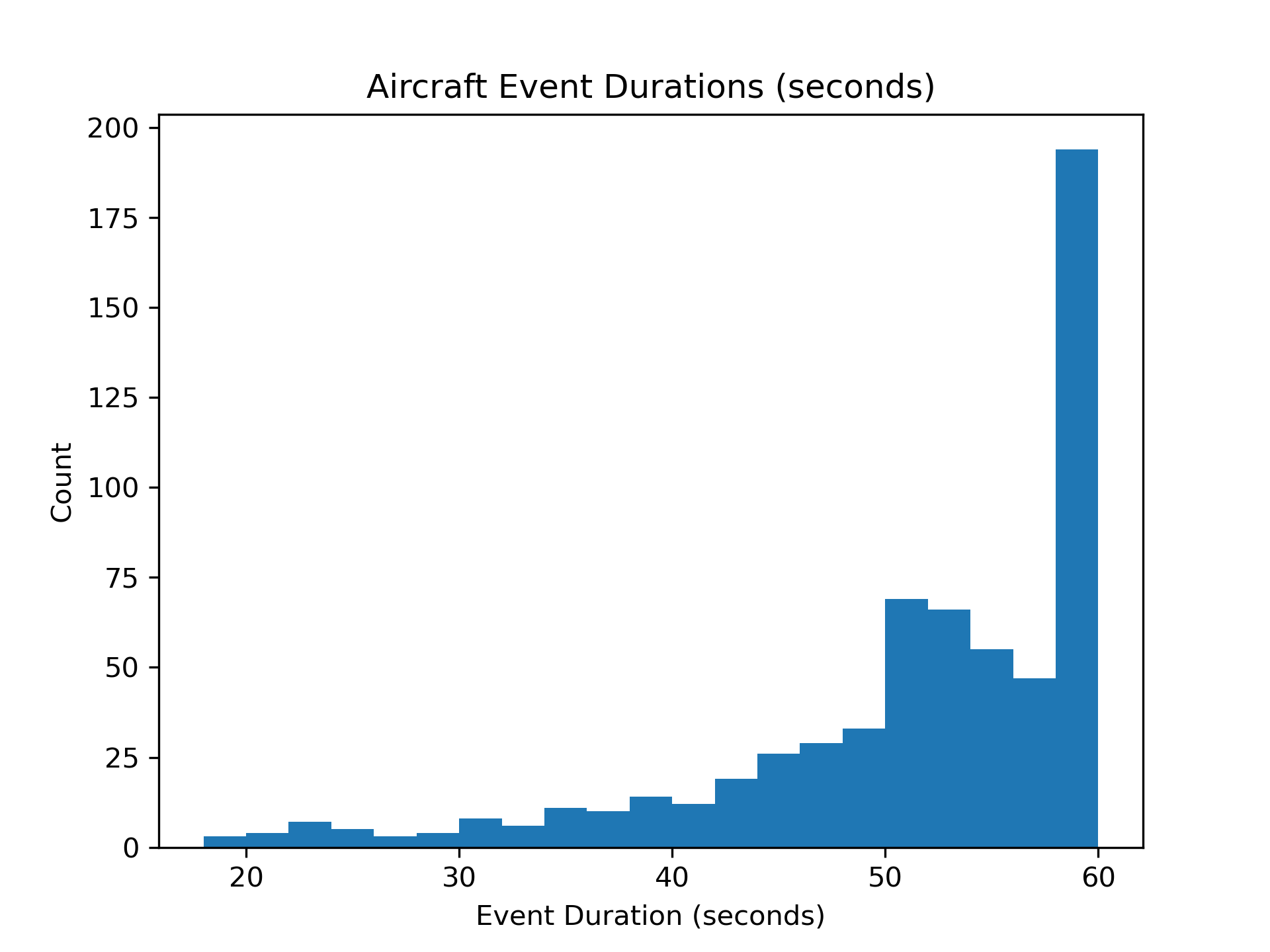}
    \caption{Histogram of aircraft audio event duration.}
    \label{fig:aircraft-sample-durations}
\end{subfigure}%
\begin{subfigure}{0.5\textwidth}
    \centering
    \includegraphics[width=0.95\linewidth, height=6cm]{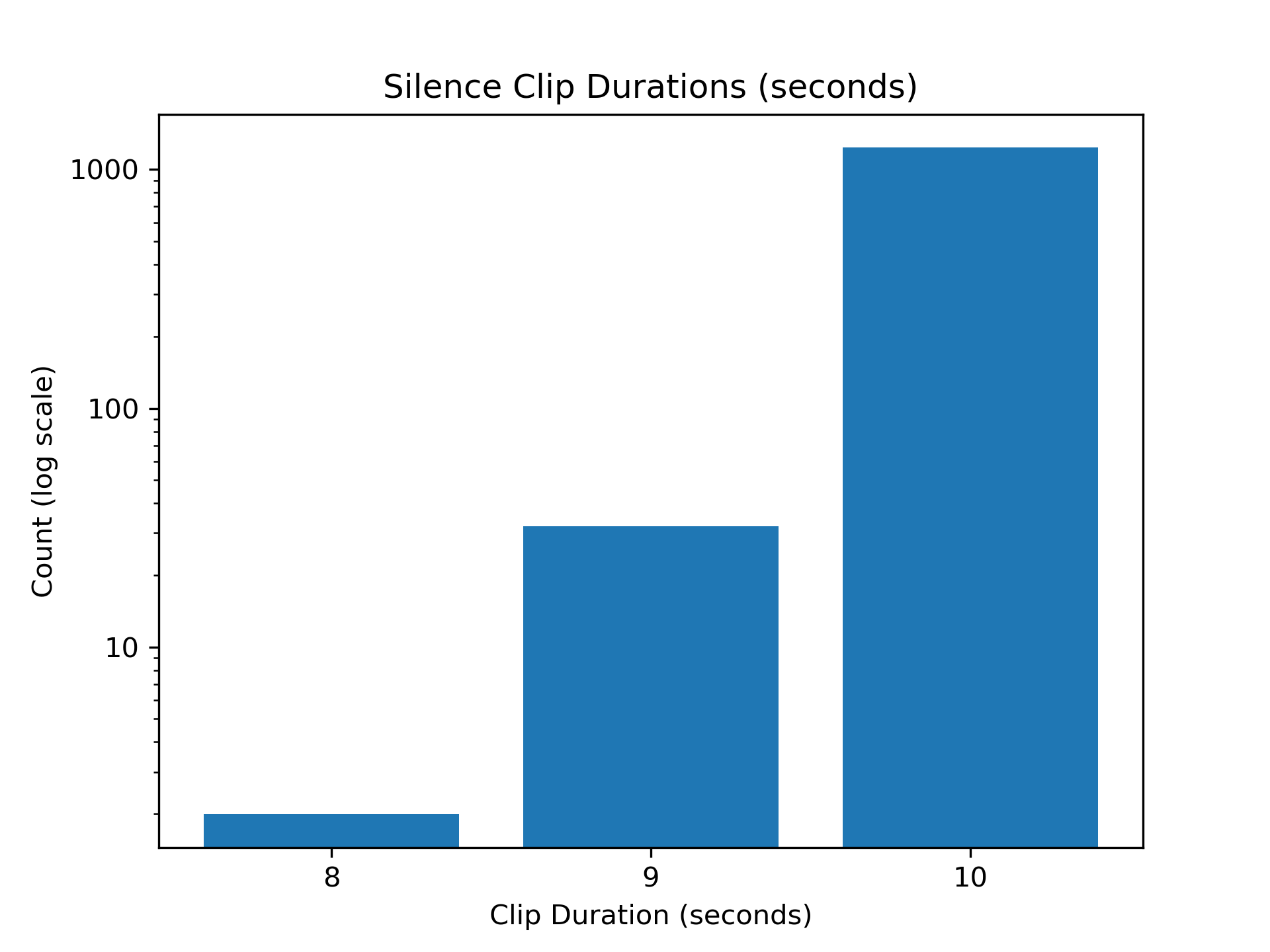}
    \caption{Histogram of silence sample duration.}
    \label{fig:silence-sample-durations}
\end{subfigure}
\caption{Duration of audio recordings.}
\label{fig:duration-distribution}
\end{figure}

The length of time an overflying aircraft is audible for varies depending on its altitude. At locations near the airport, where aircraft pass by at a low altitude - they are audible for approximately 20 seconds. On the other hand, at location "0" (where typical altitude is 3,000 feet or greater), aircraft can be audible for longer than a minute. Aircraft recordings were capped at 60 seconds to avoid an excessive dataset size, while still ensuring less salient onset and outset moments are captured. The median duration of an aircraft event was 53 seconds, and the shortest was just 18 seconds. Silence samples were capped at 10 seconds long. Some did required trimming, so the resulting silence samples ranged from 8 to 10 seconds in length. See Figure \ref{fig:duration-distribution} for distributions.

For training and inference, a segment length of 5 seconds was selected as an appropriate duration to be able to discriminate aircraft sounds from other environmental sounds (such as cars and trucks passing). Samples longer than 5 seconds were split into 5 second segments. Segments shorter than 5 seconds are zero-padded at the end to ensure consistent input feature dimensions.

\subsection{Environmental evaluation dataset}
One challenge facing this dataset is the over-representation of the positive "aircraft" class. As a means to provide a real-world scenario to evaluate model performance, an additional dataset of continuous environmental recordings was collected and annotated with aircraft events. The method of monitoring ADS-B messages described above was adapted to run parallel with a continuously recording microphone. This allowed the identification of audible aircraft throughout the environmental recordings.

The environmental evaluation dataset consists of six, one-hour long urban soundscape recordings, accompanied by a CSV file which maps class labels to their corresponding timestamp. Timestamps are quantised to five-second intervals to match the training data, resulting in 720 annotations for each hour of audio. Due to the variable length of aircraft audio events, and the lack of distinct onset and outset moments - audio segments which transition between aircraft and silence periods are tagged with an \textit{ignore} class. This is done to provide a clear boundary between silence and aircraft events, helping to avoid false misclassification at event boundaries.

\subsection{Audio feature extraction}
Librosa, a popular Python package for audio signal processing was used for transforming audio into features for training. Librosa has a high-level method for converting audio signals to Mel-band Frequency Cepstral Coefficients (MFCCs). Under the hood, this method performs a Type 2 orthonormal Discrete Cosine Transformation (DCT). 13 MFCCs were extracted per window size of 2048, with a hop length of 512 - resulting in a 216 by 13 matrix of coefficients per 5 second segment of audio.

\subsection{Evaluation}
As a means to provide a common ground for evaluating model performance, training data was split into five predefined folds for cross-validation. This method involves training the same model on five unique subsets of the training data, with the aim of eliminating potential biases arising from arbitrarily splitting the dataset. Average Precision (AP) and Precision-Recall Curves were selected as good general classification metrics for this task as they can be readily interpreted without stipulating a bias towards precision or recall. As cross-validation produces five scores per model, mean Average Precision (mAP) is reported with standard deviation to estimate overall model performance. Python package \textit{Scikit-learn}\cite{scikit-learn} was used for cross-validation and calculating scoring metrics.

\subsection{Baseline models}
Three models were selected as candidates for this binary classification task. Logistic regression (LR), multi-layer perceptron (MLP) and a convolutional neural network (CNN). A recurrent neural network with long-short term memory (RNN-LSTM) was briefly evaluated, however showed no immediate improvement over the above-mentioned models.

\subsubsection{Logistic regression}
\textit{Scikit-learn} logistic regression model with "liblinear" solver. The array of MFCCs is flattened before training. No other pre-processing or scaling is applied.

\subsubsection{Multi-layer Perceptron}
The multi-layer perceptron (MLP) is a sequential \textit{Keras}\cite{keras} model with an input layer, two hidden layers and an output layer for prediction (See Table \ref{table:mlp}). The input shape is 216 by 13 (as described above). Of the two hidden layers, the first has a size of 128 and the second with a size of 32. Both hidden layers utilise ReLU activation function, L2 regularisation factor of 0.001, and a dropout rate of 0.4 to avoid over-fitting. The final layer has a single output node with a sigmoid activation function. The Adam optimisation algorithm was selected to minimise loss (Binary cross-entropy) during training, with a learning rate of 0.001. The model was trained for 50 epochs and batch size of 216. Performance of the MLP was greatly improved by setting the \textit{class\_weights} parameter to "balanced" for training.

\subsubsection{Convolutional Neural Network}
The convolutional neural network is a sequential \textit{Keras} model with three convolutional layers, each paired with a max pooling layer and dropout. Output from the final convolutional layer is then flattened to a dense layer, which is finally fed forward to the output layer with sigmoid activation. A dropout rate of 0.4 is applied following each of the max pooling layers, and also following the penultimate dense layer. Aside from the output layer, ReLU activation function is applied to the three convolutional layers and first dense layer (See Table \ref{table:cnn}). This model was also trained for 50 epochs and with a batch size of 216. As with the MLP, CNN performance improved with the use of "balanced" class weights.

\begin{minipage}[hb]{.49\textwidth}
    \begin{table}[H]
\centering
\scriptsize
\begin{tabular}{lrr}\toprule
\addlinespace[1.2em]
&MLP \\\midrule
\addlinespace[1.2em]
INPUT &Flatten (Input shape 13 x 216) \\
\addlinespace[1.2em]
\hline
\addlinespace[1.2em]
&Dense layer (128, ReLU, L2 regularisation [0.001]) \\
&Dropout (0.4) \\
\addlinespace[1.2em]
\hline
\addlinespace[1.2em]
&Dense layer (32, ReLU, L2 regularisation [0.001]) \\
&Dropout (0.4) \\
\addlinespace[1.2em]
\hline
\addlinespace[1.2em]
OUTPUT &Dense output layer (1, Sigmoid) \\
\addlinespace[1.2em]
\bottomrule
\end{tabular}
\caption{MLP Architecture}
\label{table:mlp}
\end{table}  
\end{minipage}
\begin{minipage}[h]{.49\textwidth}
    \begin{table}[H]
\centering
\scriptsize
\begin{tabular}{lrr}\toprule
\addlinespace[0.5em]
&CNN \\\midrule
\addlinespace[0.5em]
INPUT &Input shape: (13 x 216 x 1) \\
\midrule
\addlinespace[0.5em]
&Conv2D(32, (3, 3), ReLU) \\
&MaxPooling2D((3, 3), strides=(2, 2), padding=same) \\
&Dropout (0.4) \\
\midrule
\addlinespace[0.5em]
&Conv2D(32, (3, 3), ReLU) \\
&MaxPooling2D((3, 3), strides=(2, 2), padding=same) \\
&Dropout (0.4) \\
\midrule
\addlinespace[0.5em]
&Conv2D(32, (2, 2), ReLU) \\
&MaxPooling2D((2, 2), strides=(2, 2), padding=same) \\
&Dropout (0.4) \\
\midrule
\addlinespace[0.5em]
&Flatten \\
&Dense layer (32, ReLU) \\
&Dropout (0.4) \\
\midrule
\addlinespace[0.5em]
OUTPUT &Dense output layer (1, Sigmoid) \\
\addlinespace
\bottomrule
\end{tabular}
\caption{CNN Architecture}
\label{table:cnn}
\end{table}
\end{minipage}

\section{Results}
\subsection{Dataset summary}
Samples were collected over 29 recording sessions in the period from December 8, 2022 to May 23, 2023 
(approximately 5 days of continual monitoring). The resulting dataset contains a total of 1,895 audio clips (12.39 hours) which are distributed across two top-level classes, “Aircraft” (8.87 hours/71.6\%) and “Silence” (3.52 hours/28.4\%).

\begin{figure}[H]
\centering
\begin{subfigure}{0.5\textwidth}
    \centering
    \includegraphics[width=0.9\linewidth]{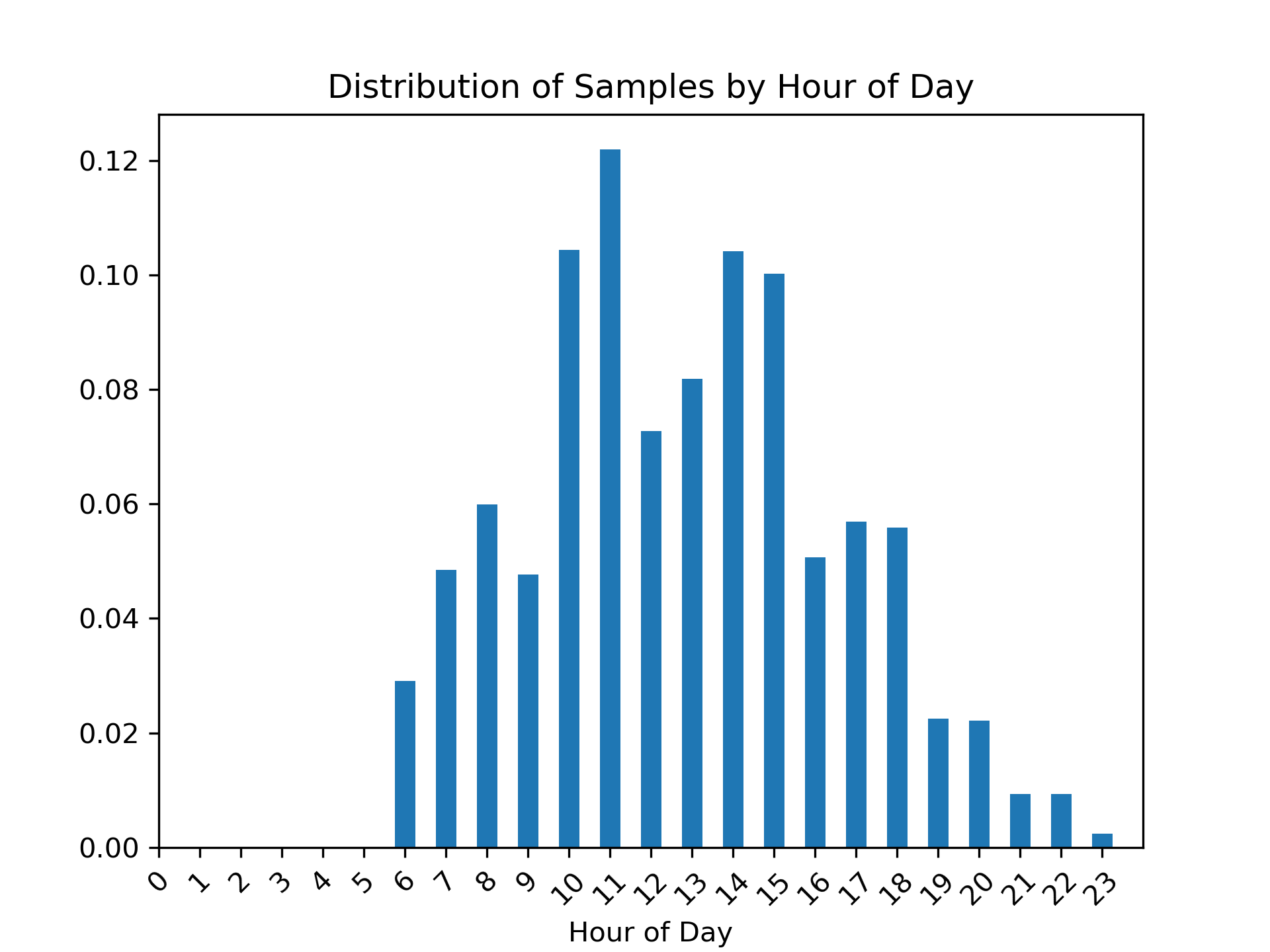}
    \caption{Distribution of recordings by hour of day}
    \label{fig:recordings-by-hour}
\end{subfigure}%
\begin{subfigure}{0.5\textwidth}
    \centering
    \includegraphics[width=0.9\linewidth]{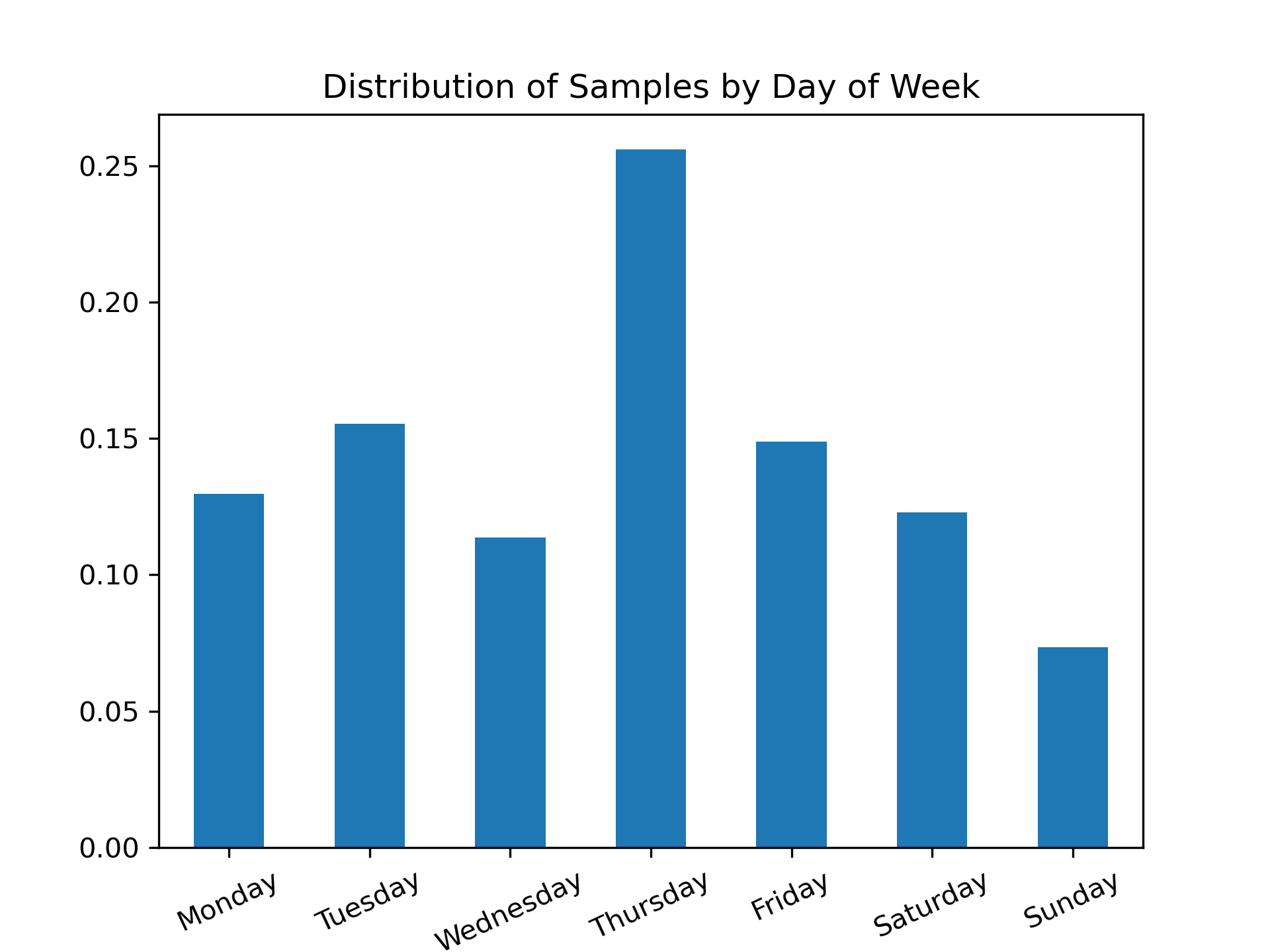}
    \caption{Distribution of recordings by day of week}
    \label{fig:recordings-by-day}
\end{subfigure}
\caption{Temporal distribution of audio recordings.}
\label{fig:temporal-distribution}
\end{figure}

The dataset was split into 6 folds - the first 5 constitute the training dataset, and the final fold is held-out as the test set for model evaluation. Table \ref{table:dataset-summary} summarises the class, location and microphone distributions across the dataset. Figures \ref{fig:recordings-by-hour} and \ref{fig:recordings-by-day} show the temporal distribution of samples.

\begin{table}[H]
\centering
\setlength{\tabcolsep}{10pt}
\begin{tabular}{lrrrrrrrr}
\toprule
 & \multicolumn{5}{c}{TRAIN/CV} & TRAIN & TEST & ENV \\
 \cmidrule(lr){2-6}\cmidrule(lr){7-7}\cmidrule(lr){8-8}\cmidrule(lr){9-9}
\textbf{fold/set} &  1  &  2  &  3  &  4  &  5  & Total  &  Total  &  Total  \\
\cmidrule(lr){1-1}\cmidrule(lr){2-6}\cmidrule(lr){7-7}\cmidrule(lr){8-8}\cmidrule(lr){9-9}
\textbf{class} &        &        &        &        &        &        &       &        \\
\cmidrule(lr){1-1}\cmidrule(lr){2-6}\cmidrule(lr){7-7}\cmidrule(lr){8-8}\cmidrule(lr){9-9}
\textbf{0 (silence)    } &  24.86 &  18.24 &  29.51 &  27.45 &  30.26 &  26.28 &  37.4 &  79.95 \\
\textbf{1 (aircraft)    } &  75.14 &  81.76 &  70.49 &  72.55 &  69.74 &  73.72 &  62.6 &  20.05 \\
\midrule
\textbf{location} &        &        &        &        &        &        &       &        \\
\cmidrule(lr){1-1}\cmidrule(lr){2-6}\cmidrule(lr){7-7}\cmidrule(lr){8-8}\cmidrule(lr){9-9}
\textbf{0    } &  100.0 &  100.0 &  94.82 &  100.0 &  95.62 &  97.98 &  92.65 &  90.87 \\
\textbf{1    } &  0.0 &  0.0 &  5.18 &  0.0 &  0.0 &  1.06 &  0.0 &  0.0 \\
\textbf{2    } &  0.0 &  0.0 &  0.0 &  0.0 &  4.38 &  0.96 &  7.35 &  9.13 \\
\midrule
\textbf{microphone} &        &        &        &        &        &        &       &        \\
\cmidrule(lr){1-1}\cmidrule(lr){2-6}\cmidrule(lr){7-7}\cmidrule(lr){8-8}\cmidrule(lr){9-9}
\textbf{0    } &  100.0 &  0.0 &  0.0 &  0.0 &  0.0 &  19.48 &  0.0 &  0.0 \\
\textbf{1    } &  0.0 &  100.0 &  100.0 &  100.0 &  100.0 &  80.52 &  100.0 &  100.0 \\
\bottomrule
\caption{AeroSonicDB sample distribution by class, location and microphone (\%).}
\label{table:dataset-summary}
\end{tabular}
\end{table}

The aircraft class includes 625 samples of 301 unique aircraft, with one particular aircraft captured 10 times (ICAO hex: 7C7777). The median altitude at time of recording is 3,250 feet, and the median MTOW (maximum take-off weight) of the recorded fleet is 73,500kg. The vast majority of aircraft samples (467 of 625 samples) are twin turbofan aeroplanes, of which Boeing 737-800's feature predominantly (231). Turboprop aircraft feature in this dataset 114 times, with the most common make and model being the Pilatus PC-12 (27). The remainder of the dataset is made up of piston-powered aeroplanes (36), helicopters (5) and quad-engine turbofan aeroplanes (3). Being a large commercial airport, these distributions fit our general expectation that commercial aircraft (typically turbofan or turboprop aeroplanes) will account for the majority of YPAD's air traffic. See Table \ref{table:airframe-features} for a detailed summary of aircraft features.

\subsection{Model evaluation}
Table \ref{table:results-table} below provides a summary of results for each of the selected models. Scores from 5-fold cross-validation are the first indication of model performance and stability. Model architecture and hyperparameters are tuned during this phase to optimise mean average precision while minimising the variance between folds. Cross-validation results with a low standard deviation suggest the hyperparameters for the model are a good fit and indicate that we should expect a similar score when evaluated against unseen data. Cross-validation results rank CNN as the best model, then MLP, followed by Logistic Regression.

\begin{table}[H]
\centering
\scriptsize
\begin{tabular}{lrrrrrrr}
    \toprule
    &\multicolumn{2}{c}{5-fold CV} &\multicolumn{2}{c}{Test Set} &\multicolumn{2}{c}{Environment Set} \\\cmidrule(lr){2-3}\cmidrule(lr){4-5}\cmidrule(lr){6-7}
    Model &Mean Ave. Precision &Std. Deviation &Mean Ave. Precision &Std. Deviation &Mean Ave. Precision &Std. Deviation \\\cmidrule(lr){1-1}\cmidrule(lr){2-3}\cmidrule(lr){4-5}\cmidrule(lr){6-7}
    Logistic Regression &98.24 &0.77 &98.64 &0.21 &84.61 &1.35 \\
    MLP &98.83 &0.86 &99.39 &0.19 &89.50 &1.50 \\
    CNN &99.14 &0.63 &98.60 &0.51 &91.32 &1.82 \\
    \bottomrule
\end{tabular}
\caption{Average precision from CV, Test and Environment evaluation.}
\label{table:results-table}
\end{table}

\subsubsection{Test set evaluation}
The test set is the first opportunity to evaluate model performance on new, unseen data. Should we observe a significant decline in model performance from cross-validation to test set evaluation, it could indicate overfitting of the model to the training data. No major changes are observed for any model against the test set. This could suggest the models generalise well to unseen data, however it could also suggest a high degree of similarity between the training and testing datasets. Ranking model performance based on the test set puts the MLP model first, followed by logistic regression, and CNN is ranked last. Judging only by the results of 5-fold cross-validation and test set evaluation, it is difficult to determine the "best" model, so further evaluation is required.

\begin{figure}[H]
\centering
\begin{subfigure}{0.33\textwidth}
    \centering
    \includegraphics[width=0.9\linewidth]{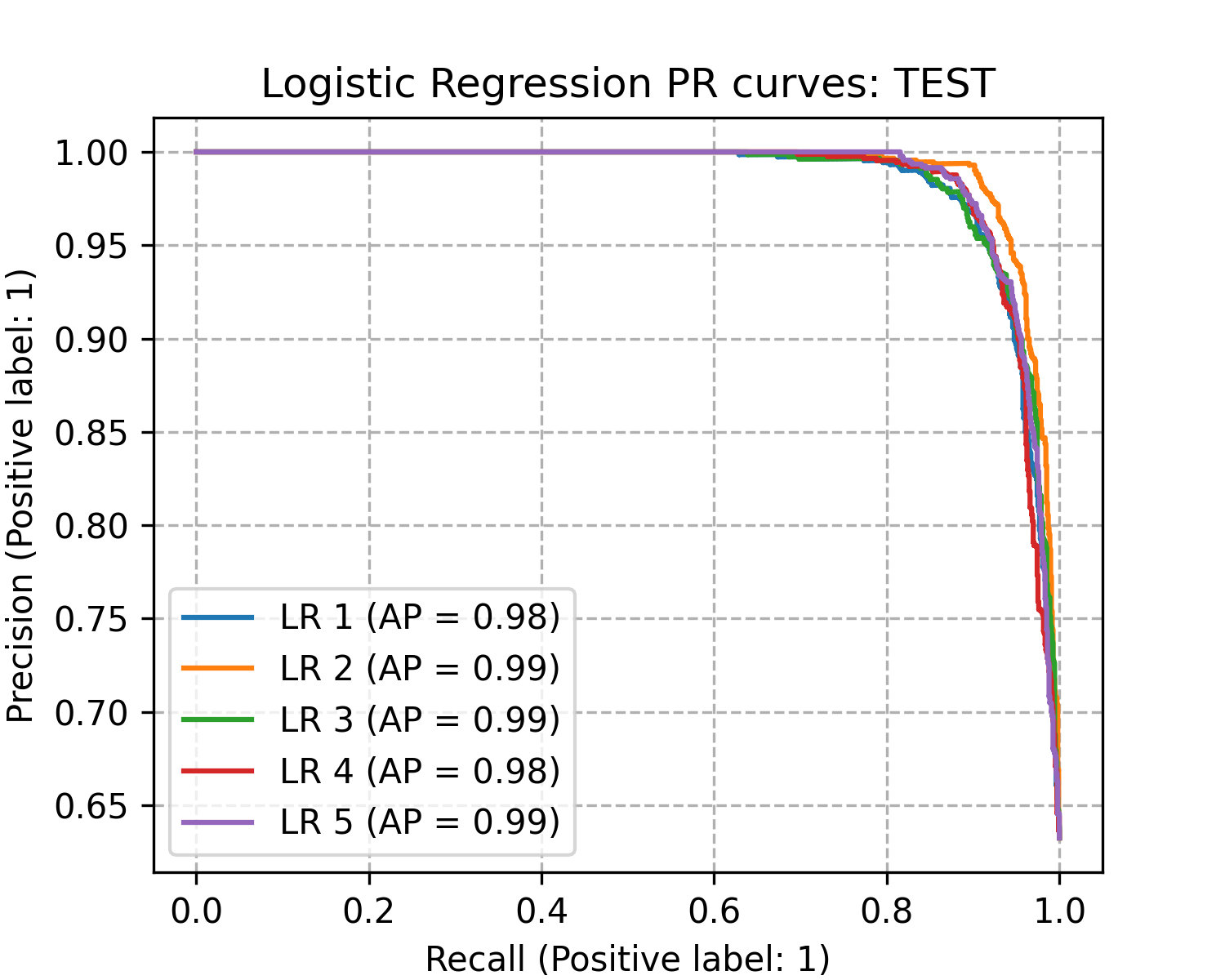}
    \caption{Logistic Regression}
    \label{fig:lr-pr-test}
\end{subfigure}%
\begin{subfigure}{0.33\textwidth}
    \centering
    \includegraphics[width=0.9\linewidth]{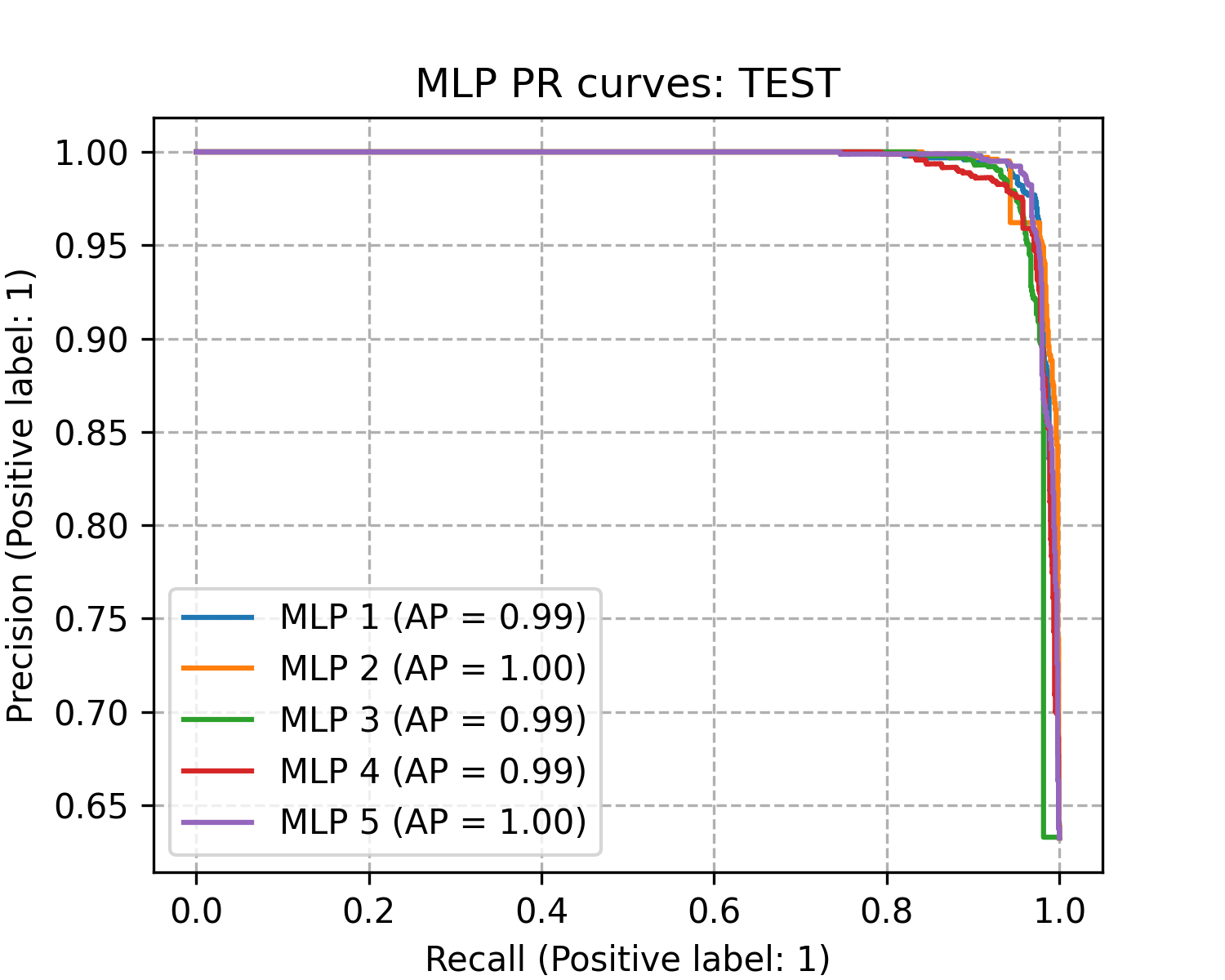}
    \caption{Multi-Layer Perceptron}
    \label{fig:mlp-pr-test}
\end{subfigure}%
\begin{subfigure}{0.33\textwidth}
    \centering
    \includegraphics[width=0.9\linewidth]{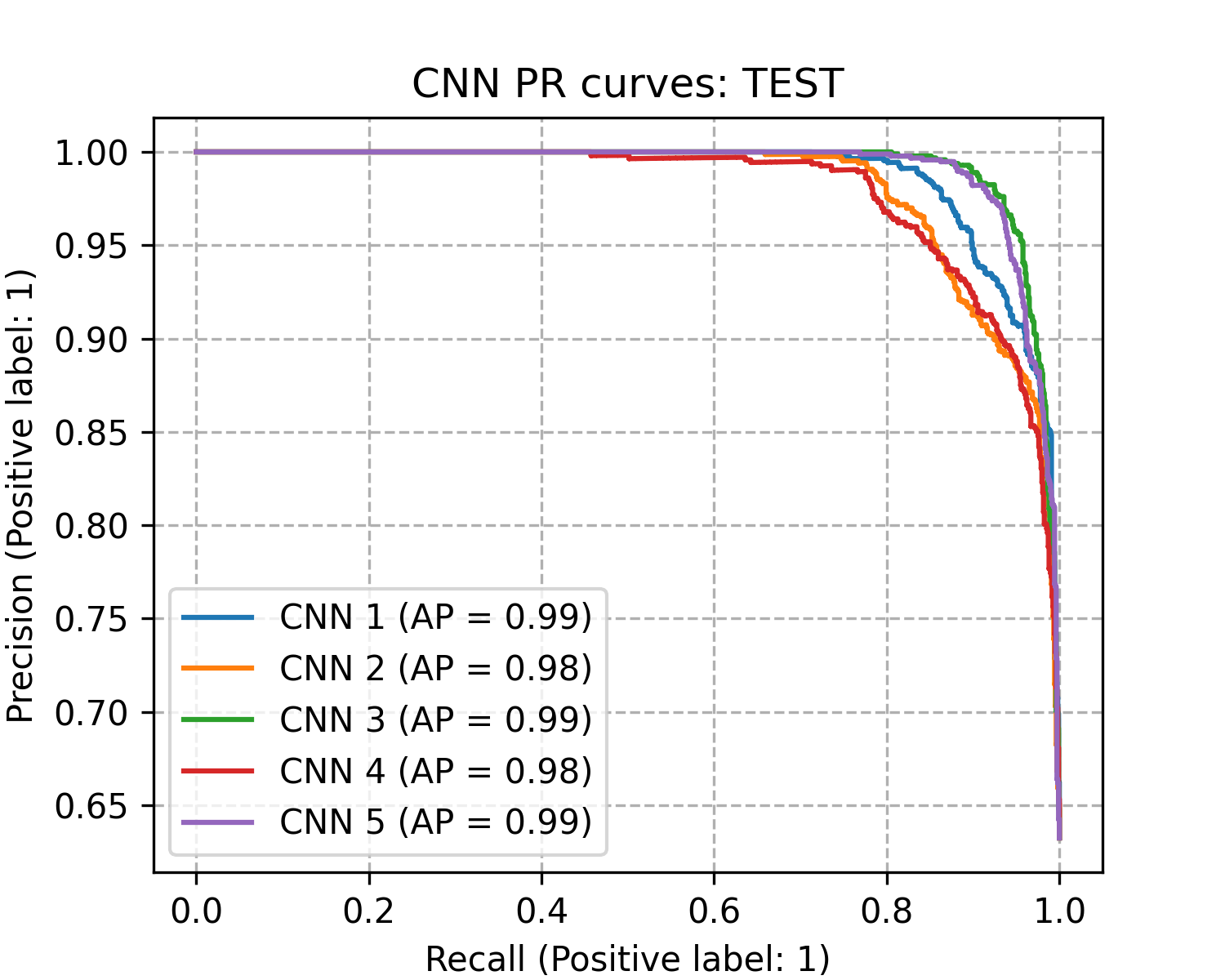}
    \caption{Convolutional Neural Network}
    \label{fig:cnn-pr-test}
\end{subfigure}
\caption{5-fold PR curves for LR, MLP and CNN models}
\label{fig:test-pr-curves}
\end{figure}

\subsubsection{Environmental evaluation}
Similarities between the training and testing data can lead to unrealistic estimates of how well a model can generalise, so the environmental evaluation set was proposed as a way to validate model performance against "difficult" real-world data. The first reason for this is the vast difference in class distribution for training and testing data compared to the "true" distribution of events. 

We found the class distributions for training and testing to be almost the inverse of what is observed in the environmental set. This adds a level of difficulty not seen in the training and testing sets, and should eliminate the chance of a model scoring well just by predicting the majority class (which is aircraft in the training data). The second advantage of the environmental set, is that it contains an aggregate of 5.28 hours of evaluation data. This is more than double the 2.35 hours in the test set, and provides a greater opportunity for distinctions to arise between models. The differences between the test and environmental sets are discussed further in section \ref{test-env-incongruities} below.

\begin{figure}[H]
\centering
\begin{subfigure}{0.33\textwidth}
    \centering
    \includegraphics[width=0.9\linewidth]{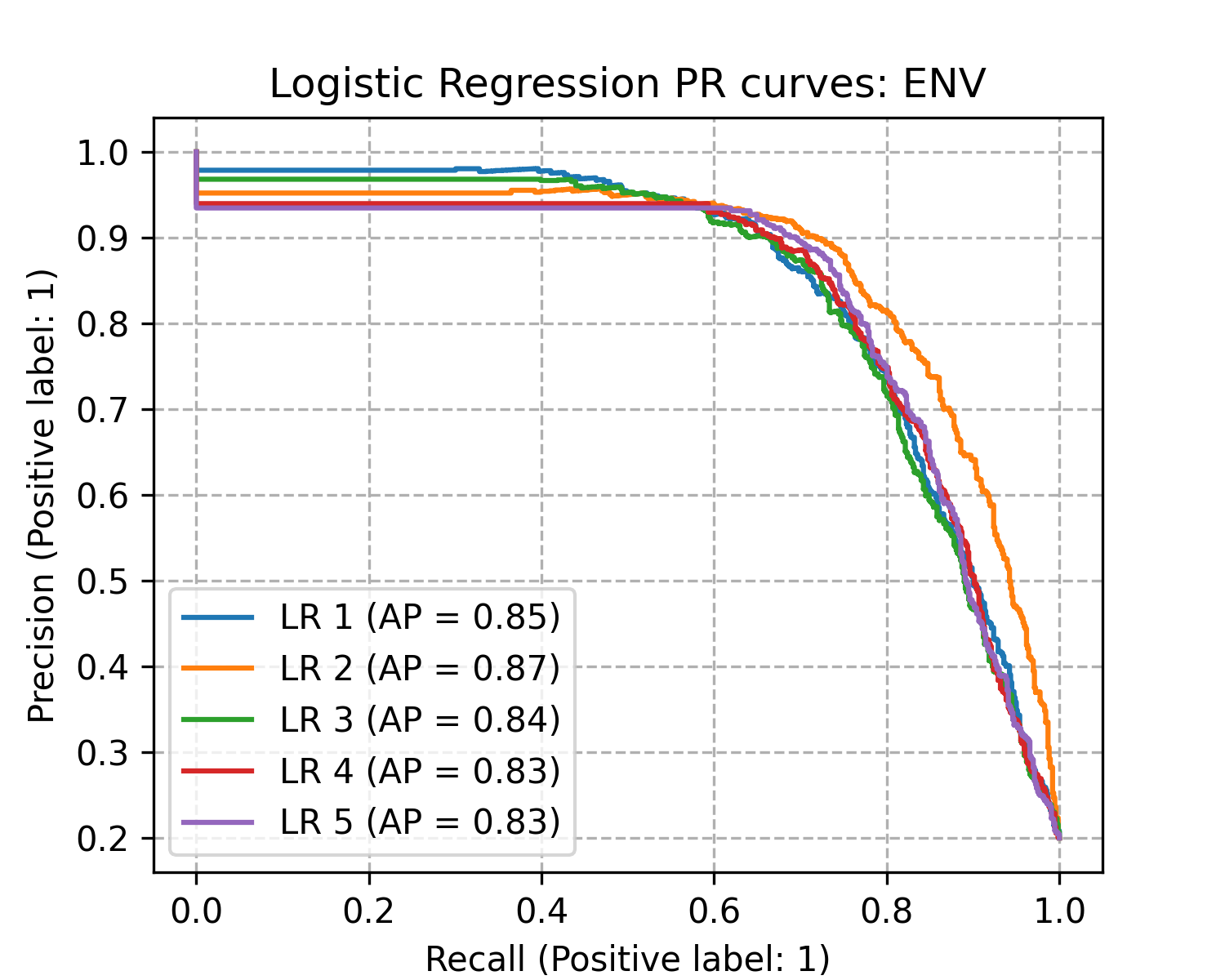}
    \caption{Logistic Regression}
    \label{fig:lr-pr-env}
\end{subfigure}%
\begin{subfigure}{0.33\textwidth}
    \centering
    \includegraphics[width=0.9\linewidth]{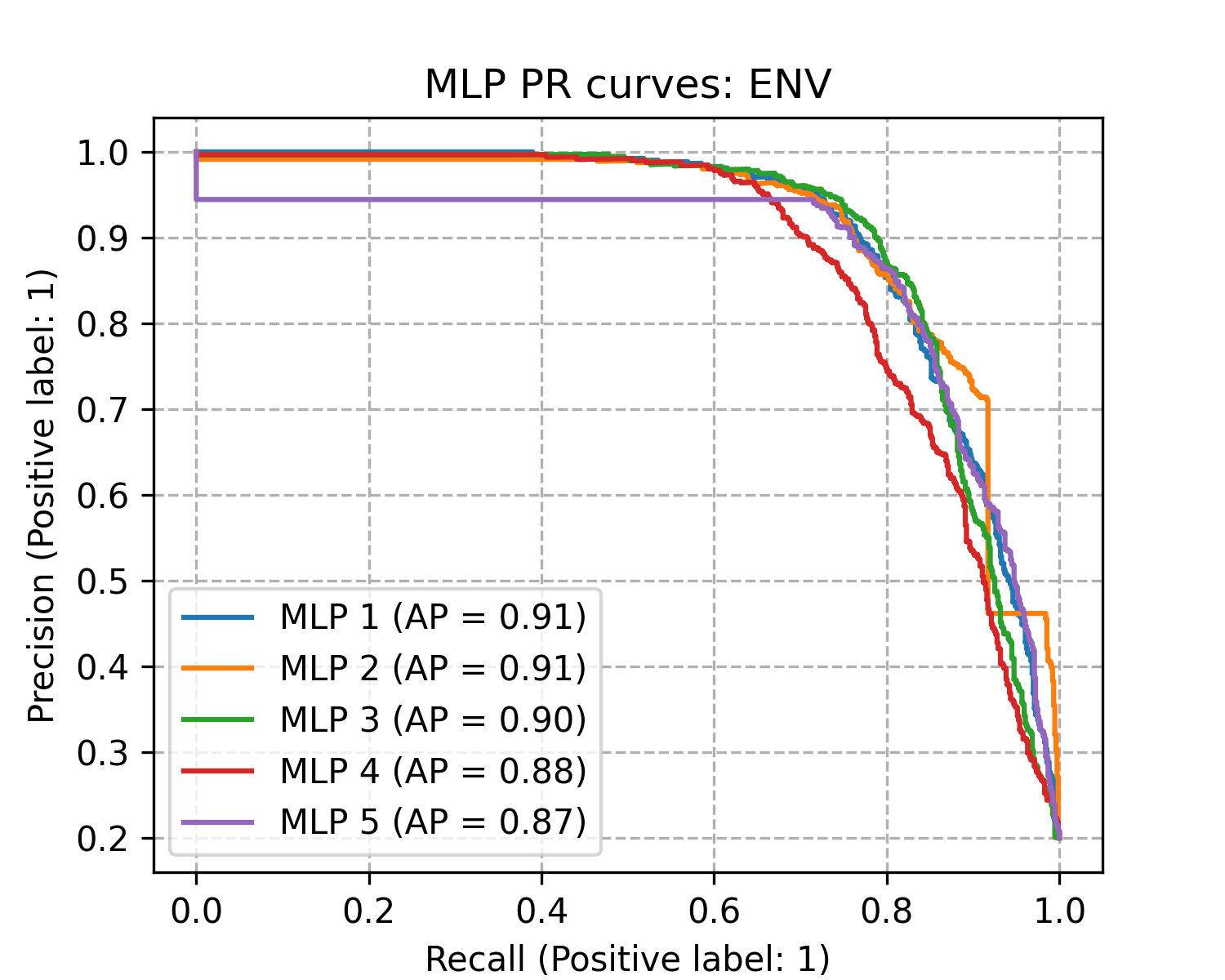}
    \caption{Multi-Layer Perceptron}
    \label{fig:mlp-pr-env}
\end{subfigure}%
\begin{subfigure}{0.33\textwidth}
    \centering
    \includegraphics[width=0.9\linewidth]{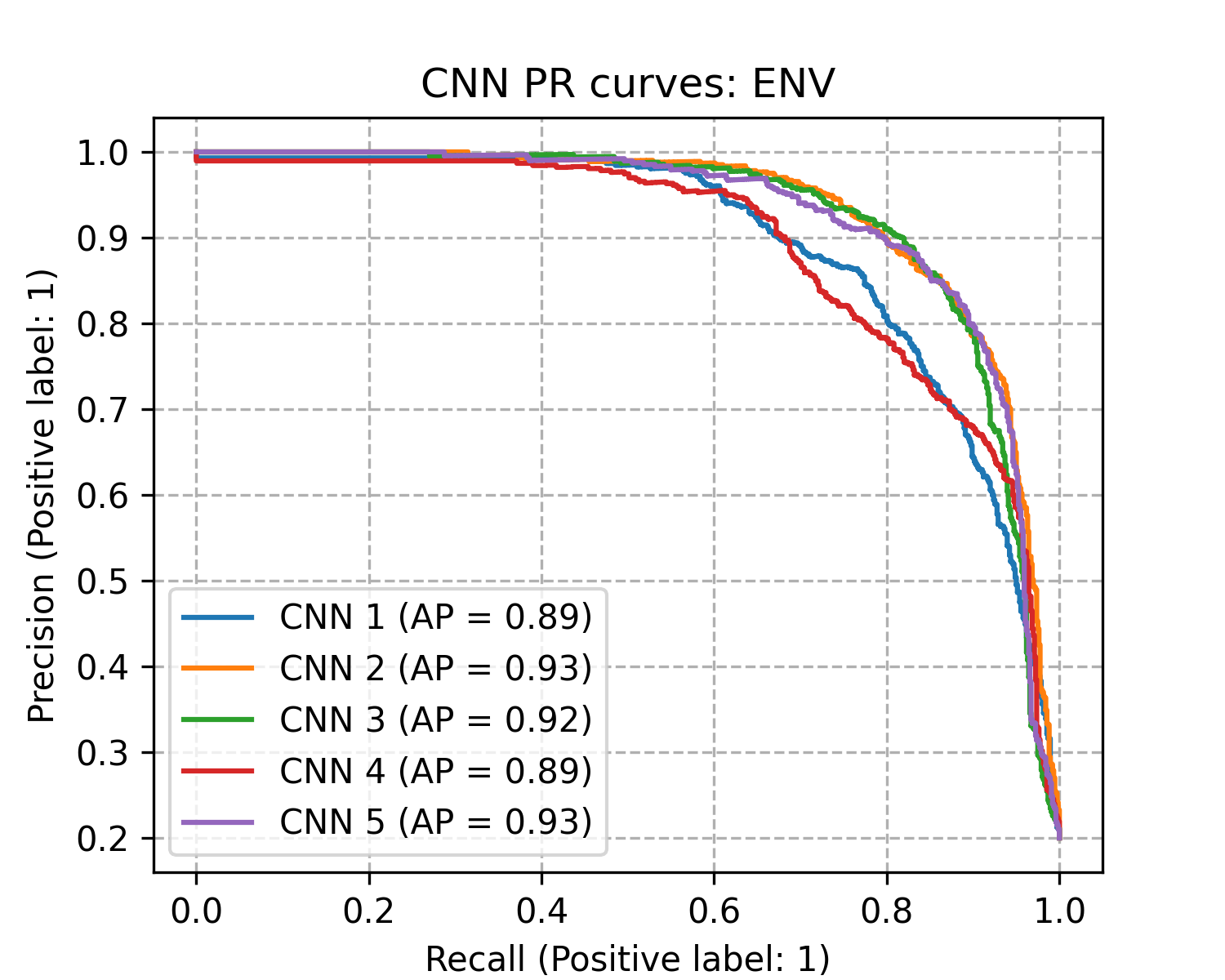}
    \caption{Convolutional Neural Network}
    \label{fig:cnn-pr-env}
\end{subfigure}
\caption{5-fold PR curves for LR, MLP and CNN models evaluated against the environment recordings}
\label{fig:env-pr-curves}
\end{figure}

After environmental evaluation, the general ability of each model starts to become more apparent. Whilst the MLP and CNN models maintain similar results to one another, the performance of logistic regression drops significantly in comparison. At this stage we can disregard logistic regression, however the choice between MLP and CNN implementations is not as clear cut. Without further fine-tuning of hyper-parameters, selecting the "best" model will require consideration of it's use-case is really a matter of considering it's use-case, and the importance of qualitative and quantitative factors available.

The environmental set offers an additional bonus for evaluation because it allows us to graphically represent how sensitive a model is to noise. To demonstrate this, each model was evaluated against each individual hour of environment (see Table \ref{table:env-results-table}). Model predictions/probabilities can then be plotted against the "ground-truth" for a more intuitive assessment of model performance.

\begin{figure}[H]
\centering
\begin{subfigure}{0.5\textwidth}
    \centering
    \includegraphics[width=0.99\linewidth]{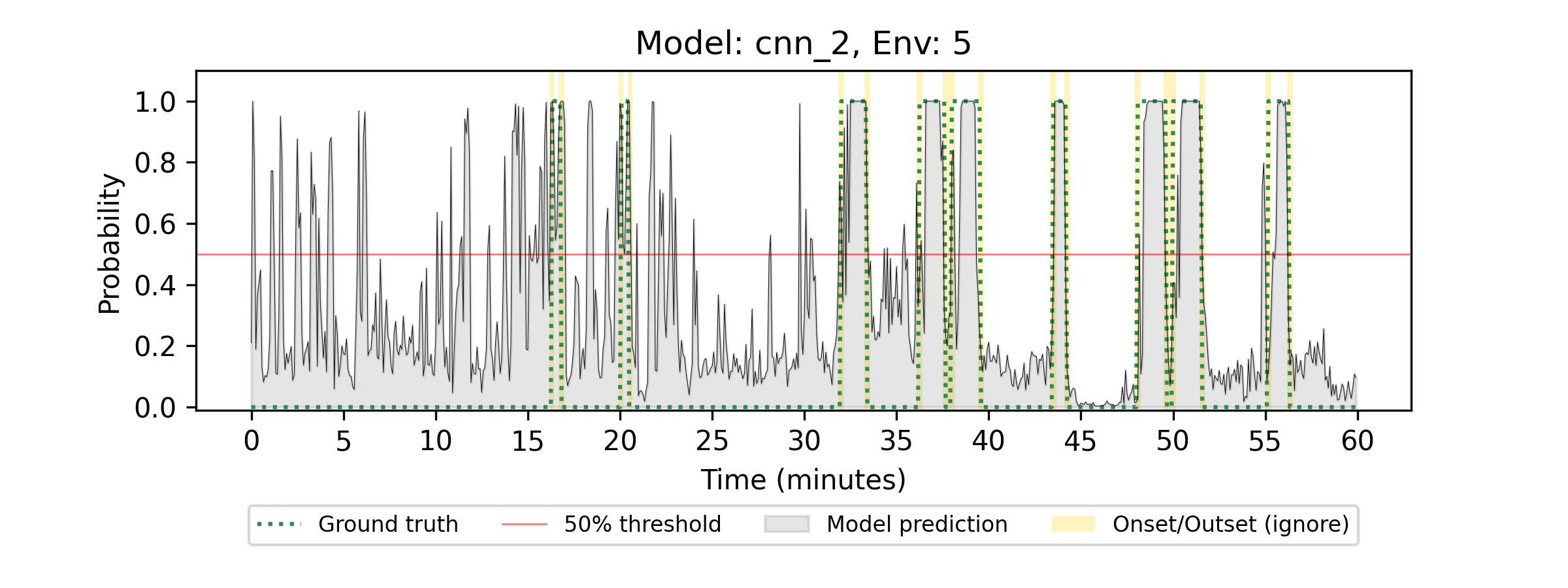}
    \caption{CNN\_2}
    \label{fig:cnn-2-hour-5}
\end{subfigure}%
\begin{subfigure}{0.5\textwidth}
    \centering
    \includegraphics[width=0.99\linewidth]{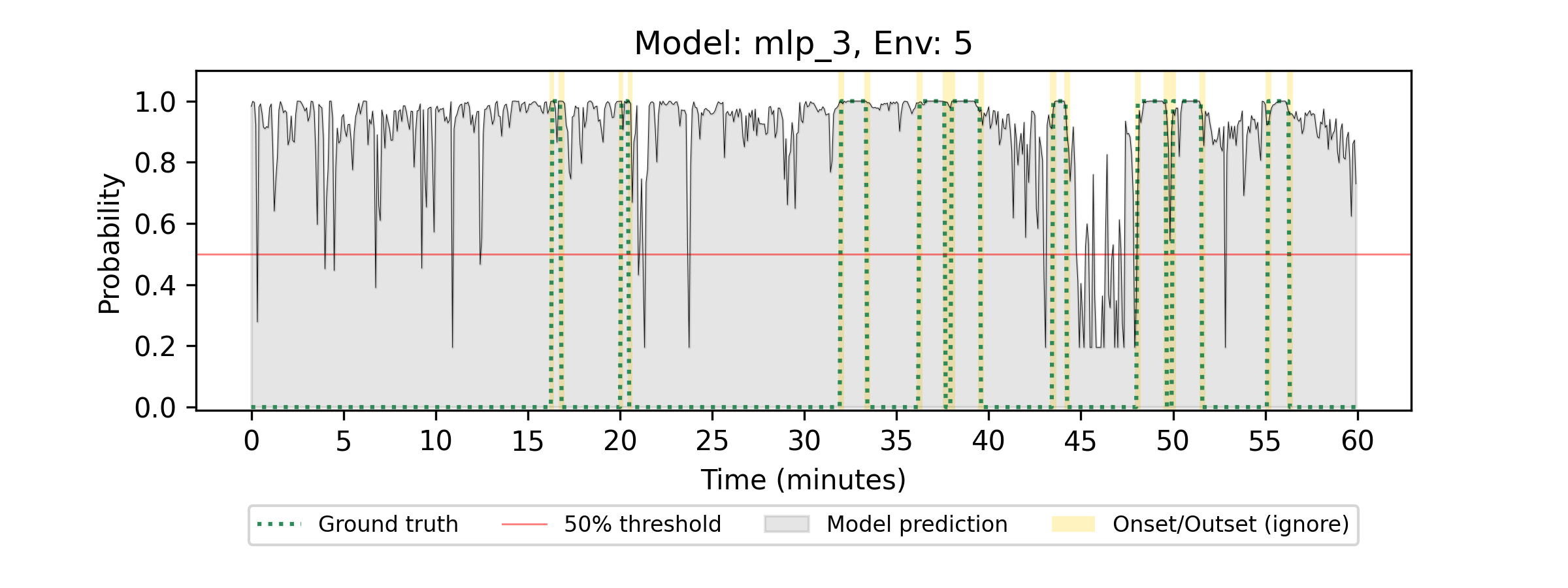}
    \caption{MLP\_3}
    \label{fig:mlp-3-hour-5}
\end{subfigure}
\caption{CNN and MLP probabilities for the fifth hour of environmental audio}
\label{fig:cnn-mlp-env-5}
\end{figure}

As mentioned above, it is difficult to choose between MLP and CNN models based solely on the quantitative results presented. For example, average precision for "MLP\_3" and "CNN\_2" is exactly the same (0.828) when evaluated against the fifth hour of environmental audio. Plotting predictions on this hour (see Figures \ref{fig:cnn-2-hour-5} and \ref{fig:mlp-3-hour-5}) does however demonstrate a clear distinction between the models. Even though these two models achieved the same AP score on this hour, the CNN is far less sensitive to noise and presents as a more robust implementation compared to the MLP. This is an interesting result, and highlights the need for a multi-faceted approach to model evaluation and selection.

\begin{table}[ht]
\addtolength{\tabcolsep}{-0.5pt}
\centering
\scriptsize
\begin{tabular}{lrrrrrrrrrrrrrrrr}
\toprule
 & \multicolumn{5}{c}{LR} & \multicolumn{5}{c}{MLP} & \multicolumn{5}{c}{CNN} &  \\
 \cmidrule(lr){2-6}\cmidrule(lr){7-11}\cmidrule(lr){12-16}
Model & \multicolumn{1}{c}{1} & \multicolumn{1}{c}{2} & \multicolumn{1}{c}{3} & \multicolumn{1}{c}{4} & \multicolumn{1}{c}{5} & \multicolumn{1}{c}{1} & \multicolumn{1}{c}{2} & \multicolumn{1}{c}{3} & \multicolumn{1}{c}{4} & \multicolumn{1}{c}{5} & \multicolumn{1}{c}{1} & \multicolumn{1}{c}{2} & \multicolumn{1}{c}{3} & \multicolumn{1}{c}{4} & \multicolumn{1}{c}{5} & Mean \\
 \cmidrule(lr){1-1}\cmidrule(lr){2-6}\cmidrule(lr){7-11}\cmidrule(lr){12-16}\cmidrule(lr){17-17}
Env &  &  &  &  &  &  &  &  &  &  &  &  &  &  &  &  \\
\midrule
1 & 0.897 & 0.955 & 0.873 & 0.911 & 0.902 & 0.921 & 0.957 & 0.915 & 0.915 & 0.927 & 0.825 & 0.901 & 0.866 & 0.849 & 0.858 & 0.898 \\
2 & 0.929 & 0.950 & 0.910 & 0.939 & 0.930 & 0.989 & 0.981 & 0.985 & 0.980 & 0.987 & 0.986 & 0.993 & 0.983 & 0.996 & 0.978 & \cellcolor{green!18}0.968 \\
3 & 0.940 & 0.964 & 0.925 & 0.885 & 0.945 & 0.957 & 0.970 & 0.958 & 0.931 & 0.953 & 0.980 & 0.990 & 0.986 & 0.983 & 0.984 & 0.957 \\
4 & 0.861 & 0.898 & 0.833 & 0.863 & 0.822 & 0.927 & 0.934 & 0.920 & 0.923 & 0.934 & 0.923 & 0.948 & 0.921 & 0.949 & 0.938 & 0.906 \\
5 & 0.680 & 0.618 & 0.730 & 0.628 & 0.644 & 0.819 & 0.770 & 0.828 & 0.746 & 0.615 & 0.736 & 0.828 & 0.868 & 0.654 & 0.869 & \cellcolor{red!30}0.736 \\
6 & 0.882 & 0.926 & 0.889 & 0.883 & 0.885 & 0.949 & 0.945 & 0.939 & 0.926 & 0.948 & 0.951 & 0.949 & 0.947 & 0.950 & 0.950 & 0.928 \\
\midrule
Mean & 0.865 & 0.885 & 0.860 & \cellcolor{red!30}0.852 & 0.855 & 0.927 & 0.926 & 0.924 & 0.904 & 0.894 & 0.900 & \cellcolor{green!18}0.935 & 0.929 & 0.897 & 0.930 & 0.899 \\
\bottomrule
\end{tabular}
\caption{Average precision per model per hour of environmental evaluation set.}
\label{table:env-results-table}
\end{table}

\section{Conclusion}

\begin{figure}[H]
\centering
\includegraphics[width=0.95\textwidth]{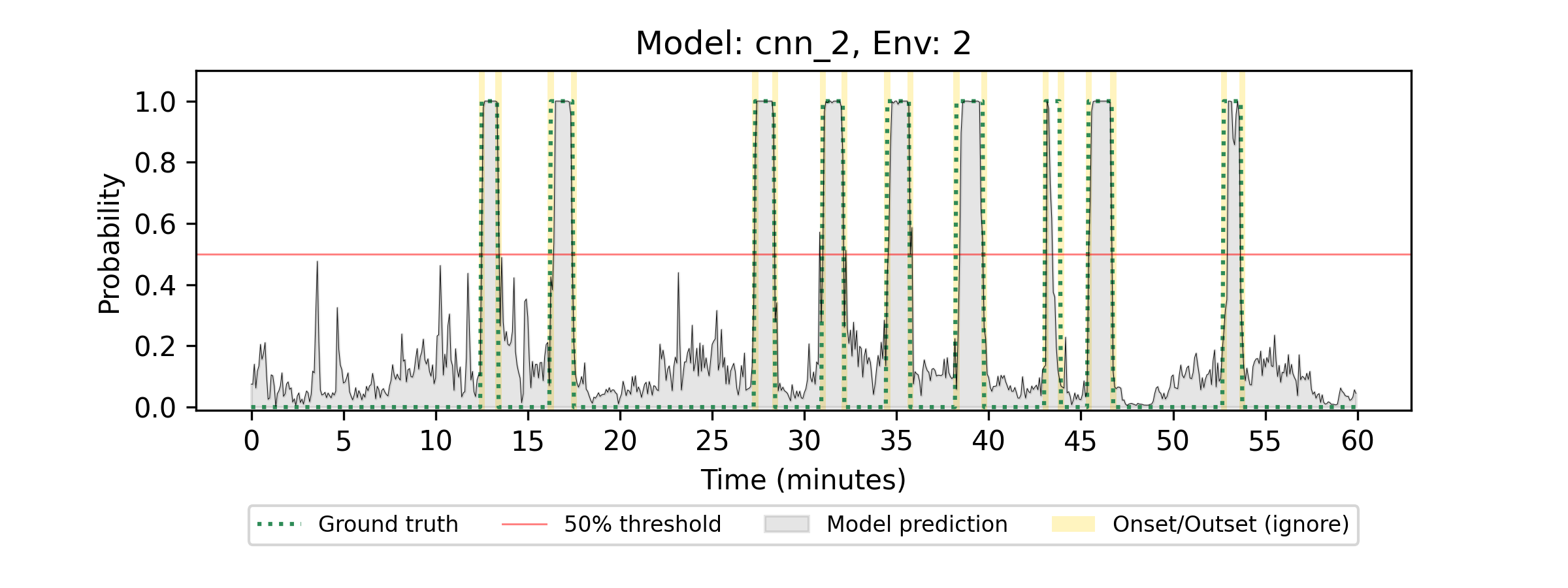}
\caption{(CNN\_2) probabilities for the "easiest" hour of environmental audio.}
\label{fig:cnn-2-env-2}
\end{figure}

Considering the above, the Convolutional Neural Network stands out as a good benchmark for acoustic aircraft detection. The CNN performed very well in the environmental evaluations at location "0" (where 98\% of the training data was collected), and  seemed to generalise the best when presented with difficult edge-cases (environmental hour 5). These results imply that less than five days of monitoring and recording can capture enough audio to train a robust location specific binary classifier. Figure \ref{fig:cnn-2-env-2} demonstrates the "best" overall model (CNN\_2) performs across the "easiest" hour (2) of environmental recordings.

\section{Discussion}
\subsection{Incongruities between Test and Environmental evaluation sets} \label{test-env-incongruities}
Table \ref{table:results-table} shows how the results for each model do not vary significantly between cross-validation and test evaluation. However, we do observe a significant drop in model performance across the board when evaluated against the environmental set. The training and test sets were collected with the same methodology, and they share very similar characteristics with respect to class and location distributions. The training set for example has roughly 74\% aircraft to 26\% silence, with 98\% from location "0" and 1\% from locations "1" and "2" respectively. Similarly, the test set contains approximately 63\% aircraft to 37\% silence, with 93\% from location "0" and 7\% from location "2". In contrast, the class distribution of the environmental evaluation set is around 20\% aircraft to 80\% silence. This is significantly different from the training data, however should be considered a more realistic estimate of the true distribution. 

Another major contributing factor in the differences between the evaluation sets is location. Just 2\% of training data came from locations nearby the airport, whereas 9\% of the environmental evaluation data was collected from location "2" (the first 30 minutes of environment hour 5). These 30 minutes predominantly feature close proximity road-traffic, and only two aircraft events lasting around 30 seconds each. The difficulty of the fifth hour can been seen in Table \ref{table:env-results-table}, where all models have consistently under-performed in this hour. Upon inspection of model predictions against the ground truth (see Figure \ref{fig:cnn-2-hour-5}), it is obvious this location is a point of confusion for even the best model. This highlights the importance of curating an evaluation dataset which can truly tests model performance against difficult edge cases and unseen data.

\subsection{Incongruities between Environmental evaluation results}
Results in Table \ref{table:results-table} and Figure \ref{fig:env-pr-curves} above do vary slightly from the averages presented in Table \ref{table:env-results-table}. This is due to the scores presented in Table \ref{table:results-table} and Figure \ref{fig:env-pr-curves} being based on the aggregated environment set, whereas Table \ref{table:env-results-table} first evaluates against each hour of environment before aggregating the results.

\section{Further work}

Given the lack of helicopter (5 samples) and piston-engine aircraft (36 samples) samples in this version of the dataset, robust classification for these sub-classes of aircraft may not be feasible with this data alone. Future versions should aim to fill these gaps to provide a more comprehensive training corpus for machine listening tasks.

The collection of additional ADS-B messages such as; heading, speed, vertical speed, latitude and longitude will greatly broaden the applications for this dataset. Heading for example could be used to determine if the aircraft is taking-off or landing. Latitude and longitude can be combined with altitude to calculate approximate distance from the recording device. Synchronisation of tracking data (altitude, speed, heading, distance) with the audio sample could be utilised to develop acoustic detection and ranging models.

Finally, the development of computationally lightweight models that can be deployed on micro-controllers to perform real-time inference. Such a system could be paired with a sound-level meter to create an inexpensive aircraft noise monitoring unit which doesn't rely on external data streams, such as internet, airport radar or ADS-B data. These units could readily be deployed to investigate noise at a particular location, or spread across a city as an array for testing the validity of modelled airport noise contours. 

\section*{ACKNOWLEDGEMENTS}

Valerio Velardo and The Sound of AI Community


\clearpage
\bibliographystyle{ieeetr}
\bibliography{bibliography.bib}

\begin{thebibliography}{10}

\bibitem{downward_blake_2023_8371595}
B.~Downward, ``Aerosonicdb (ypad-0523): Labelled audio dataset for acoustic detection and classification of aircraft,'' Sept. 2023.

\bibitem{World_Health_Organisation_Regional_Office_for_Europe2019-za}
{World Health Organisation: Regional Office for Europe}, ``{ENVIRONMENTAL} {NOISE} {GUIDELINES} for the european region.'' \url{https://www.who.int/europe/publications/i/item/9789289053563}, Jan. 2019.

\bibitem{ISO20906}
I.~O. for Standardization, ``{Acoustics — Unattended monitoring of aircraft sound in the vicinity of airports},'' standard, International Organization for Standardization, Geneva, CH, Dec. 2009.

\bibitem{adelaide_nfpms}
A.~Australia, ``The adelaide noise and flight path monitoring system (nfpms).'' \url{https://aircraftnoise.airservicesaustralia.com/2022/11/11/the-adelaide-noise-and-flight-path-monitoring-system-nfpms/}, Nov. 2022.
\newblock Accessed: 2023-08-27.

\bibitem{noise_fp_monitoring}
L.~C. Kenna, ``Noise and flight path monitoring at australian airports,'' in {\em 2004 Australian Acoustical Society Conference on Acoustics}, pp.~491--494, 2004.

\bibitem{airservices_noise_monitoring}
A.~Australia, ``Noise and flight path monitoring system.'' \url{https://www.airservicesaustralia.com/community/environment/aircraft-noise/monitoring-aircraft-noise/}, May 2023.
\newblock Accessed: 2023-08-27.

\bibitem{casa_ads-b_def}
C.~A. S.~A. (Australia), ``Ads-b subsidy deadline extended.'' \url{https://www.casa.gov.au/ads-b-subsidy-deadline-extended}, May 2023.
\newblock Accessed: 2023-07-22.

\bibitem{casa_cao_20.18}
C.~A. S.~A. (Australia), ``Civil aviation order 20.18.'' \url{https://www.legislation.gov.au/Details/F2017C01115}, Nov. 2017.
\newblock Accessed: 2023-07-19.

\bibitem{casa_vfr_ads-b}
C.~A. S.~A. (Australia), ``New standards for automatic dependent surveillance – broadcast (ads-b) equipment for vfr aircraft - (cd 1905as).'' \url{https://consultation.casa.gov.au/regulatory-program/cd-1905as/}, June 2020.
\newblock Accessed: 2023-07-19.

\bibitem{cartwright2020sonyc}
M.~Cartwright, J.~Cramer, A.~E.~M. Mendez, Y.~Wang, H.-H. Wu, V.~Lostanlen, M.~Fuentes, G.~Dove, C.~Mydlarz, J.~Salamon, {\em et~al.}, ``Sonyc-ust-v2: An urban sound tagging dataset with spatiotemporal context,'' {\em arXiv preprint arXiv:2009.05188}, 2020.

\bibitem{salamon_dataset_2014:urbansound8k}
J.~Salamon, C.~Jacoby, and J.~P. Bello, ``A dataset and taxonomy for urban sound research,'' in {\em Proceedings of the 22nd {ACM} international conference on Multimedia}, {MM} '14, pp.~1041--1044, Association for Computing Machinery, 2014.

\bibitem{piczak2015dataset:esc50}
K.~J. Piczak, ``Esc: Dataset for environmental sound classification,'' in {\em Proceedings of the 23rd ACM International Conference on Multimedia}, MM '15, (New York, NY, USA), p.~1015–1018, Association for Computing Machinery, 2015.

\bibitem{audio_set}
J.~F. Gemmeke, D.~P.~W. Ellis, D.~Freedman, A.~Jansen, W.~Lawrence, R.~C. Moore, M.~Plakal, and M.~Ritter, ``Audio set: An ontology and human-labeled dataset for audio events,'' in {\em 2017 IEEE International Conference on Acoustics, Speech and Signal Processing (ICASSP)}, pp.~776--780, 2017.

\bibitem{nordby2021automatic}
J.~Nordby, F.~Nemazi, and D.~Rieber, ``Automatic detection of noise events at shooting range using machine learning,'' 2021.

\bibitem{nasa-aircraft-noise-prediction}
W.~E. Zorumski, ``Aircraft noise prediction program theoretical manual.,'' tech. rep., National Aeronautics and Space Administration, 1982.

\bibitem{shi2011detecting}
W.~Shi, G.~Arabadjis, B.~Bishop, P.~Hill, R.~Plasse, and J.~Yoder, ``Detecting, tracking, and identifying airborne threats with netted sensor fence,'' {\em Sensor Fusion-Foundation and Applications}, pp.~139--158, 2011.

\bibitem{aircraft-detection-2009}
R.~O. Nielsen, ``Acoustic detection of low flying aircraft,'' in {\em 2009 IEEE Conference on Technologies for Homeland Security}, pp.~101--106, 2009.

\bibitem{acoustic-aircraft-tracking-2013}
A.~Sutin, H.~Salloum, A.~Sedunov, and N.~Sedunov, ``Acoustic detection, tracking and classification of low flying aircraft,'' in {\em 2013 IEEE International Conference on Technologies for Homeland Security (HST)}, pp.~141--146, 2013.

\bibitem{drone-dataset-2022}
Y.~Wang, Z.~Chu, I.~Ku, E.~C. Smith, and E.~T. Matson, ``A large-scale uav audio dataset and audio-based uav classification using cnn,'' in {\em 2022 Sixth IEEE International Conference on Robotic Computing (IRC)}, pp.~186--189, 2022.

\bibitem{drone-detection-2021}
S.~Al-Emadi, A.~Al-Ali, and A.~Al-Ali, ``Audio-based drone detection and identification using deep learning techniques with dataset enhancement through generative adversarial networks,'' {\em Sensors}, vol.~21, no.~15, 2021.

\bibitem{c-uas-2021}
J.~Wang, Y.~Liu, and H.~Song, ``Counter-unmanned aircraft system(s) (c-uas): State of the art, challenges, and future trends,'' {\em IEEE Aerospace and Electronic Systems Magazine}, vol.~36, no.~3, pp.~4--29, 2021.

\bibitem{giladi2020real}
R.~Giladi, ``Real-time identification of aircraft sound events,'' {\em Transportation Research Part D: Transport and Environment}, vol.~87, p.~102527, 2020.

\bibitem{nooelec_sdr}
Nooelec, ``Nooelec nesdr mini 2+ 0.5ppm tcxo usb rtl-sdr receiver (rtl2832 + r820t2).'' \url{https://www.nooelec.com/store/nesdr-mini-2-plus.html}, July 2023.
\newblock Accessed: 2023-07-22.

\bibitem{dump_1090}
S.~Sanfilippo and M.~Robb, ``Dump1090 mode s decoder.'' \url{https://github.com/MalcolmRobb/dump1090}, Oct. 2014.

\bibitem{adsb-exchange}
A.-B. Exchange, ``World's largest source of unfiltered flight data.'' \url{https://www.adsbexchange.com/}.

\bibitem{radarbox}
A.~Systems, ``Radarbox offers flight data such as latitude and longitude positions, origins and destinations, flight numbers, aircraft types, altitudes, headings and speeds..'' \url{https://radarbox.com/}.

\bibitem{airport-data}
A.~Data, ``One-stop aviation information - includes airport location, runway information, radio navigation aids, communication frequencies, nearby hotels, maps, aircraft registration database, airport photo and airplane photo gallery..'' \url{https://www.airport-data.com/}.

\bibitem{plane-spotters}
P.~Spotters, ``Aviation photos, airline fleets \& more | planespotters.net.'' \url{https://www.planespotters.net/}.

\bibitem{casa-aircraft-register}
CASA, ``Data files for registered aircraft..'' \url{https://services.casa.gov.au/CSV/acrftreg.csv}, May 2023.
\newblock Accessed: 2023-05-06.

\bibitem{scikit-learn}
F.~Pedregosa, G.~Varoquaux, A.~Gramfort, V.~Michel, B.~Thirion, O.~Grisel, M.~Blondel, P.~Prettenhofer, R.~Weiss, V.~Dubourg, J.~Vanderplas, A.~Passos, D.~Cournapeau, M.~Brucher, M.~Perrot, and E.~Duchesnay, ``Scikit-learn: Machine learning in {P}ython,'' {\em Journal of Machine Learning Research}, vol.~12, pp.~2825--2830, 2011.

\bibitem{keras}
F.~Chollet {\em et~al.}, ``Keras,'' 2015.

\end{thebibliography}

\clearpage
\section{Appendix}
\begin{table}[!htbp]\centering
\caption{Summary of aircraft features.}\label{table:airframe-features}
\scriptsize
\setlength{\arrayrulewidth}{0.05mm}
\setlength{\tabcolsep}{10pt}
\begin{tabular}{lrrrrr}\toprule
\textbf{feature} &\textbf{description} &\textbf{n unique values} &\textbf{n most frequent} &\textbf{n samples} \\\midrule
\multirow{5}{*}{\textit{hex\_id}} &\multirow{5}{*}{Unique ICAO 24-bit address} &\multirow{5}{*}{301} &7C7777 &10 \\
& & &7C68CE &9 \\
& & &7C779A &8 \\
& & &7C2BEE &8 \\
& & &7C6C52 &7 \\
\hline
\multirow{3}{*}{\textit{altitude}} &\multirow{3}{*}{Approximate altitude in feet} &\multirow{3}{*}{89} &min &-225 \\
& & &median &3,250 \\
& & &max &5,675 \\
\hline
\multirow{2}{*}{\textit{airframe}} &\multirow{2}{*}{Mechanical structure of the aircraft} &\multirow{2}{*}{2} &Power Driven Aeroplane &620 \\
& & &Rotorcraft &5 \\
\hline
\multirow{4}{*}{\textit{engtype}} &\multirow{4}{*}{Type of engine} &\multirow{4}{*}{4} &Turbofan &470 \\
& & &Turboprop &114 \\
& & &Piston &37 \\
& & &Turboshaft &4 \\
\hline
\multirow{3}{*}{\textit{engnum}} &\multirow{3}{*}{Number of engines} &\multirow{3}{*}{3} &2 &583 \\
& & &1 &39 \\
& & &4 &3 \\
\hline
\multirow{5}{*}{\textit{shortdesc}} &\multirow{5}{*}{Short code describing the airframe and engine} &\multirow{5}{*}{8} &L2J &467 \\
& & &L2T &86 \\
& & &L1T &28 \\
& & &L2P &26 \\
& & &L1P &10 \\
\hline
\multirow{5}{*}{\textit{typedesig}} &\multirow{5}{*}{ICAO type designator for make and model} &\multirow{5}{*}{41} &B738 &231 \\
& & &E190 &88 \\
& & &A320 &68 \\
& & &PC12 &27 \\
& & &F100 &26 \\
\hline
\multirow{5}{*}{\textit{manu}} &\multirow{5}{*}{Aircraft manufaturer } &\multirow{5}{*}{22} &THE BOEING COMPANY &256 \\
& & &EMBRAER &89 \\
& & &AIRBUS INDUSTRIE &83 \\
& & &BOMBARDIER INC &35 \\
& & &PILATUS AIRCRAFT LTD &33 \\
\hline
\multirow{5}{*}{\textit{model}} &\multirow{5}{*}{Aircraft model} &\multirow{5}{*}{57} &737-838 &104 \\
& & &737-8FE &99 \\
& & &ERJ 190-100 IGW &77 \\
& & &A320-232 &68 \\
& & &PC-12/47E &27 \\
\hline
\multirow{5}{*}{\textit{engmanu}} &\multirow{5}{*}{Engine manufacturer} &\multirow{5}{*}{10} &CFM INTERNATIONAL, S.A. &244 \\
& & &GENERAL ELECTRIC COMPANY &128 \\
& & &PRATT \& WHITNEY CANADA &94 \\
& & &INTERNATIONAL AERO ENGINES &68 \\
& & &TEXTRON LYCOMING &37 \\
\hline
\multirow{5}{*}{\textit{engmodel}} &\multirow{5}{*}{Engine model} &\multirow{5}{*}{61} &V2527-A5 &68 \\
& & &CFM-56-7B24 &59 \\
& & &CFM56-7B26/3 &47 \\
& & &CF34-10E5 &42 \\
& & &CF34-10E6 &41 \\
\hline
\multirow{5}{*}{\textit{engfamily}*} &\multirow{5}{*}{Engine family - *derived from "engmodel"} &\multirow{5}{*}{27} &CFM56 &242 \\
& & &CF34 &88 \\
& & &V2500 &68 \\
& & &PT6 &48 \\
& & &PW100 &35 \\
\hline
\multirow{2}{*}{\textit{fueltype}} &\multirow{2}{*}{Fuel type} &\multirow{2}{*}{2} &Kerosene &588 \\
& & &Gasoline &37 \\
\hline
\multirow{5}{*}{\textit{propmanu}} &\multirow{5}{*}{Propeller manufacturer} &\multirow{5}{*}{6} &NOT FITTED WITH PROPELLER &475 \\
& & &HARTZELL PROPELLERS &53 \\
& & &DOWTY PROPELLERS &32 \\
& & &HAMILTON STANDARD &29 \\
& & &MT PROPELLERS &21 \\
\hline
\multirow{5}{*}{\textit{propmodel}} &\multirow{5}{*}{Propeller model} &\multirow{5}{*}{35} &NOT APPLICABLE &475 \\
& & &14SF-15 &16 \\
& & &R408/6-123-F/17 &13 \\
& & &HC-E5A-3A/NC10245B &11 \\
& & &HC-D4N-3A &9 \\
\hline
\multirow{3}{*}{\textit{mtow}} &\multirow{3}{*}{Maximum take off weight (MTOW) in kilograms} &\multirow{3}{*}{58} &min &1,134 \\
& & &median &73,500 \\
& & &max &351,800 \\
\bottomrule
\end{tabular}
\end{table}

\end{document}